# Summing Divergent Perturbative Series in a Strong Coupling Limit. The Gell-Mann–Low Function of the $\varphi^4$ Theory

**I. M. Suslov**

*Kapitza Institute of Physical Problems, Russian Academy of Sciences, Moscow, 117334 Russia*
*e-mail: suslov@kapitza.ras.ru*

**Abstract**—An algorithm is proposed for determining asymptotics of the sum of a perturbative series in the strong coupling limit using given values of the expansion coefficients. Application of the algorithm is illustrated, methods for estimating errors are developed, and an optimization procedure is described. Applied to the $\varphi^4$ theory, the algorithm yields the Gell-Mann–Low function asymptotics of the type $\beta(g) \approx 7.4g^{0.96}$ for large $g$. The fact that the exponent is close to unity can be interpreted as a manifestation of the logarithmic branching of the type $\beta(g) \sim g(\ln g)^{-\gamma}$ (with $\gamma \approx 0.14$), which is confirmed by independent evidence. In any case, the $\varphi^4$ theory is self-consistent. The procedure of summing perturbative series with arbitrary values of the expansion parameter is discussed.

## 1. INTRODUCTION

This paper presents a systematic description of the algorithm proposed previously in a brief communication [1]. Operation of the algorithm is illustrated by test examples, methods for estimating errors are developed, and an optimization procedure is described. Using this algorithm, the Gell-Mann–Low function of the $\varphi^4$ theory—the main physical result of this study—can be reconstructed with a tenfold greater precision.

The abstract formulation of the problem is as follows. Let some function $W(g)$ be expanded into a series of the perturbation theory in powers of a coupling constant $g$:

$$W(g) = \sum_{N=0}^{\infty} W_N(-g)^N. \tag{1}$$

The first several expansion coefficients $W_N$ can be obtained by straightforward diagram calculations. The high-order terms can be determined using the Lipatov method [2], which is applicable to most of the important problems and yields for $W_N$ an asymptotic behavior of the type (see reviews [3–5]):

$$W_N^{as} = ca^N\Gamma(N+b) \approx ca^N N^{b-1} N!. \tag{2}$$

Matching asymptotics (2) to the first coefficients provides information about all terms of the series and allows the $W(g)$ function to be approximately restored, but this procedure requires using special methods for summing divergent series. Implementation of this approach allowed the critical indices of the phase transition theory to be determined to within the third decimal position [6–8], thus rendering the intermediate coupling region ($g \sim 1$) principally accessible. However, this direction was not developed further because the problem of renormalon contributions arose that cast doubt [9] on the applicability of the Lipatov method. The interest in this field had dropped sharply and no breakthrough into the strong coupling region took place.

Expanding the theory into the strong coupling region is required in many fields of theoretical physics. The most known cases, related to the dependence of the effective coupling constant $g$ on the spatial scale $L$, include the problem of electrodynamics at very small distances and the confinement problem. The dependence of $g$ on $L$ in renormalizable theories is determined by the equation

$$-\frac{dg}{d\ln L^2} = \beta(g) = \beta_2 g^2 - \beta_3 g^3 + \beta_4 g^4 - \dots. \tag{3}$$

In the general case, this description requires information on the Gell-Mann–Low function $\beta(g)$ for arbitrary $g$. The possible variants were classified by Bogolyubov and Shirkov [10]. In the case of $\beta_2 > 0$, the situation reduces to the following. If the function $\beta(g)$ possesses a root at $g_0$, then $g(L) \longrightarrow g_0$ as $L \longrightarrow 0$. If $\beta(g)$ at large $g$ behaves as $g^\alpha$ with $\alpha \le 1$, then $g(L) \longrightarrow \infty$ at small $L$; should $\beta(g)$ grow as $g^\alpha$ with $\alpha > 1$, the theory is no longer self-consistent and cannot describe the behavior of $g(L)$ in the entire range of $L$.

The first attempt at restoring the β function in the $\varphi^4$ theory with the Euclidean action

$$S\{\varphi\} = \int d^4x\left\{\frac{1}{2}(\partial\varphi)^2 + \frac{16\pi^2}{4!}g\varphi^4\right\} \quad (4)$$

was undertaken by Popov *et al.* [11]. The Shirkov group attempted to move into the strong coupling region [12] and obtained for large $g$ asymptotics of the type $0.9g^2$, which differs only by a coefficient from a one-loop law $1.5g^2$ valid for $g \longrightarrow 0$. A close asymptotic behavior ($1.06g^{1.9}$) was obtained by Kubyshin [13], while the more recently developed variational perturbation theory of Sissakian *et al.* [14] yields $2.99g^{1.5}$. All these results give evidence that the $\varphi^4$ theory is not self-consistent.[1] This is, however, rather strange from the standpoint of condensed-matter applications, where a quite reasonable disordered system model [16, 17] well defined in the continuum limit is mathematically strictly reduced to the $\varphi^4$ theory. Another argument follows from the author's recent study [9] showing the $\varphi^4$ theory to contain no renormalon singularities, which can be considered as evidence of self-consistency. This situation makes revision of the above results an urgent task.

In this paper, an algorithm is proposed for restoring asymptotics of the sum of a perturbative series in the strong coupling limit using given values of the expansion coefficients (Section 2). Application of the algorithm is illustrated by test examples with both known expansion coefficients (Section 4) and the coefficients obtained by interpolation (Sections 5 and 6). Methods for estimating errors and an optimization procedure are developed (Sections 3 and 6). The problem of summing the perturbative series with finite $g$ is considered, and it is demonstrated that knowledge of the $W(g)$ asymptotics significantly increases precision of the results (Section 7). The main physical result of this study consists in reconstructing the Gell-Mann–Low function of the $\varphi^4$ theory (Section 8). The task is solved proceeding from the same information as that used in [13], namely, the first four coefficients of expansion of the $\beta(g)$ function in the subtraction scheme [15, 18]

$$\beta(g) = \frac{3}{2}g^2 - \frac{17}{6}g^3 + \frac{154.14}{8}g^4 - \frac{2338}{16}g^5 + \ldots, \quad (5)$$

and their asymptotics according to Lipatov, taking into account the first-order correction [19]:

$$\beta_N = \frac{1.096}{16\pi^2}N^{7/2}N!\left\{1 - \frac{4.7}{N} + \ldots\right\}. \quad (6)$$

[1] It should be noted that Kazakov *et al.* [12] do not insist on this conclusion, emphasizing the preliminary character of their results (see also [15]).

Note that the interaction term in expression (4) corresponds to the "natural" charge normalization, for which the parameter $a$ in asymptotics (2) is unity. It will be demonstrated that the results obtained in [12, 13] are not artifacts: they objectively reflect the behavior $\beta(g)$ in the interval $1 \lesssim g \lesssim 10$. However, the true asymptotics is manifested at still greater $g$ and gives evidence of self-consistency of the $\varphi^4$ theory.

## 2. RELATIONSHIP BETWEEN $W(g)$ ASYMPTOTICS AND EXPANSION COEFFICIENTS

Let us formulate the problem of restoring the asymptotics

$$W(g) = W_\infty g^\alpha, \quad g \longrightarrow \infty, \quad (7)$$

using the coefficients $W_N$ of the series (1). This coefficients, increasing at large $N$ according to the factorial law (2), are assumed to be set numerically. By analogy with the case of critical indices introduced in the phase transition theory, the slow (logarithmic) corrections to (7) are considered as overstating the accuracy. For exponentially growing $W(g)$, which can be revealed by abnormally large values of α, the series (1) is considered upon preliminarily taking the logarithm.

### *2.1. Standard (Conform-Borel) Summing Procedure*

Considering the sum of series (2) in the Borel sense [20], we use a modified definition of the Borel image $B(g)$,

$$W(g) = \int_0^\infty dx e^{-x}x^{b_0-1}B(gx),$$
$$B(g) = \sum_{N=0}^\infty B_N(-g)^N, \quad B_N = \frac{W_N}{\Gamma(N+b_0)}, \quad (8)$$

where $b_0$ is an arbitrary parameter (convenient for optimization of the summation procedure [6]). It was suggested by Le Guillou and Zinn-Justin [6] and recently proved for the $\varphi^4$ by the author [9] that the Borel image is analytical in the complex plane $g$ cut from $-1/a$ to $-\infty$ (Fig. 1a). The analytical continuation of $B(g)$ from the convergence circle $|g| < 1/a$ to an arbitrary complex $g$ value is provided by a conformal mapping $g = f(u)$ of the plane with a cut into a unity circle $|u| < 1$ (Fig. 1b). The re-expansion of $B(g)$ into a series in $u$,

$$B(g) = \sum_{N=0}^\infty B_N(-g)^N\Big|_{g=f(u)} \longrightarrow B(u) = \sum_{N=0}^\infty U_N u^N, \quad (9)$$

gives a series converging for any $g$. Indeed, all the possible singular points ($P$, $Q$, $R$, …) of the $B(g)$ function occur on the cut and their images ($P$, $Q$, $Q'$, $R$, $R'$, …) fall on the boundary $|u| = 1$ of the circle. Therefore, the second series in (9) converges at any $u < 1$, but the interior of this circle is in a single-valued correspondence with the region of analyticity in the $g$ plane.

The conformal mapping is defined by the formulas

$$g = \frac{4}{a}\frac{u}{(1-u)^2} \quad \text{or} \quad u = \frac{(1+ag)^{1/2}-1}{(1+ag)^{1/2}+1}, \qquad (10)$$

from which we readily find a relationship between $U_N$ and $B_N$:

$$U_0 = B_0, \quad U_N = \sum_{K=1}^{N} B_K\left(-\frac{4}{a}\right)^K C_{N+K-1}^{N-K} \qquad (11)$$
$$(N \geq 1).$$

In order to establish a relationship between asymptotics (7) and the expansion coefficients, we will use the fact that the behavior of $U_N$ at large $N$ is determined by a sum of the contributions from singular points occurring on the boundary $|u| = 1$. This can be readily checked by expressing $U_N$ in terms of $B(u)$,

$$U_N = \oint_C \frac{du}{2\pi i}\frac{B(u)}{u^{N+1}}, \qquad (12)$$

and deforming the integration contour (enclosing the point $u = 0$) so as to make it passing around the cuts from all singular points to infinity. A singularity of the type $A(1 - u/u_0)^\beta$ at the point $u_0 = e^{i\varphi}$ makes a contribution to $U_N$ of the type

$$\frac{A}{\Gamma(-\beta)}\frac{e^{-i\varphi N}}{N^{1+\beta}}. \qquad (13)$$

Now we can readily find the contributions to $U_N$ from the singular points of the initial Borel image $B(g)$. For power singularities at the points $g = \infty$, $g = -1/a$, and $g = g_0$ with $g_0 \in (-\infty, -1/a)$, the corresponding expressions are as follows:

$$B(g) = Ag^\alpha \longrightarrow U_N = \frac{A}{\Gamma(2\alpha)}\left(\frac{4}{a}\right)^\alpha \frac{1}{N^{1-2\alpha}},$$

$$B(g) = A(g+1/a)^\beta \longrightarrow U_N = \frac{A}{(4a)^\beta\Gamma(-2\beta)}\frac{(-1)^N}{N^{1+2\beta}},$$
$$B(g) = A(g-g_0)^\beta \longrightarrow U_N \qquad (14)$$
$$= \frac{2A}{\Gamma(-\beta)}\left(\frac{\cos(\varphi/2)}{a\sin^3(\varphi/2)}\right)^\beta \frac{\cos(\varphi N - \pi\beta/2)}{N^{1+\beta}},$$

where $\varphi = \arccos(1 + 2/ag_0)$.

The singularities of $B(g)$ change depending on the parameter $b_0$ in formulas (8). For the Borel images $B(g)$

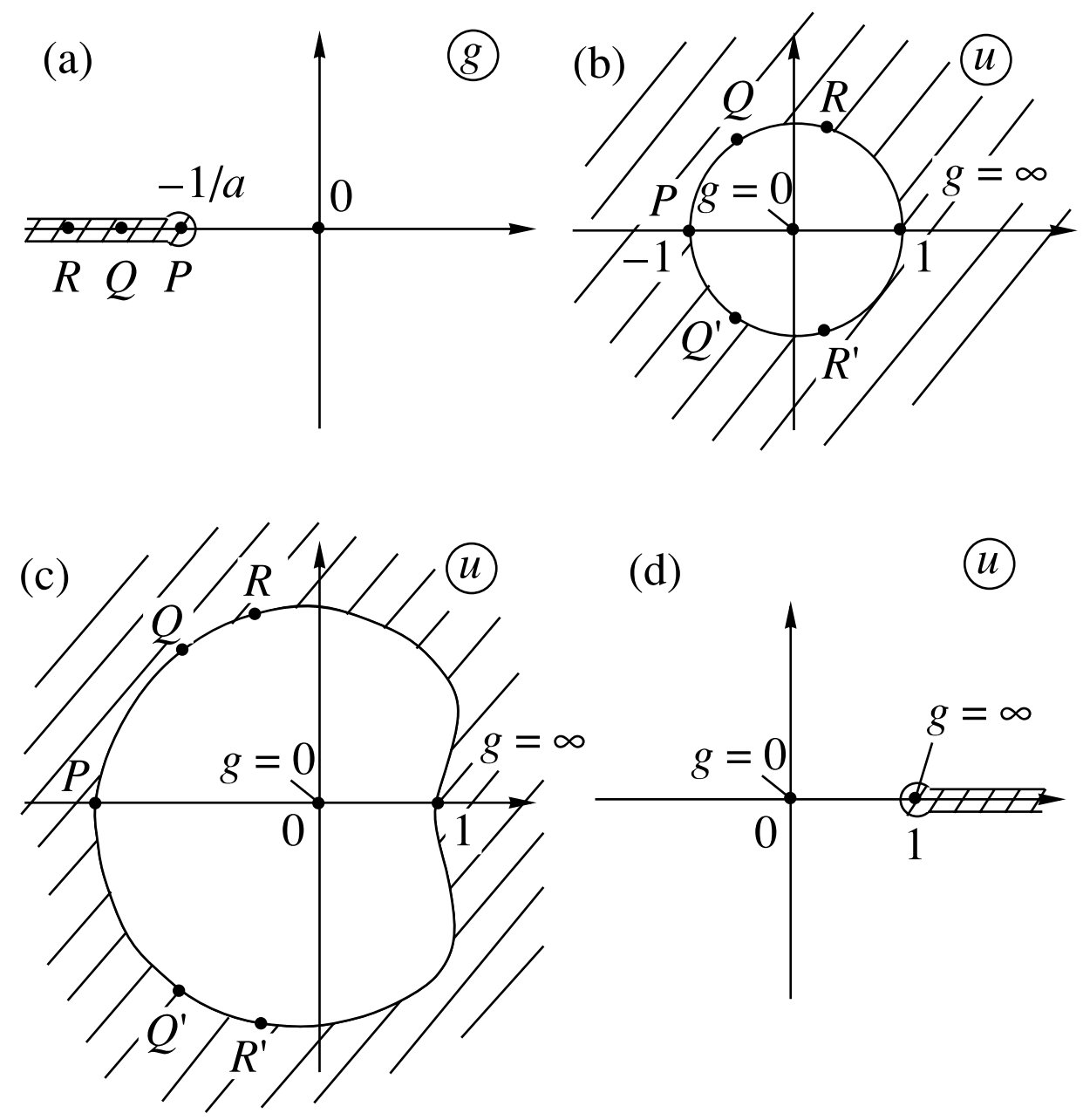

**Fig. 1.** (a) The Borel image is analytical in the complex plane with $(-\infty, -1/a)$ cut; (b) this analyticity region can be conformally mapped onto the unity circle; (c) restricting the consideration to analytical continuation to the positive semiaxis, the conformal mapping is admitted onto any region in which the point $u = 1$ is the closest boundary point to the origin; (d) in the extremal form (18) of this mapping, the analyticity region can be conformally mapped onto the plane with $(1, \infty)$ cut.

and $\tilde{B}(g)$ corresponding to $b_0$ and $b_1$, we readily obtain a recalculation formula

$$\tilde{B}(g) = \frac{1}{\Gamma(b_1-b_0)}\int_0^\infty dx \frac{x^{b_1-b_0-1}}{(1+x)^{b_1}} B\left(\frac{g}{1+x}\right), \qquad (15)$$

and a rule of singularity transformation at a finite ($g_0$) or infinite points on the passage from $b_0$ to $b_1$:

$$B(g) = A\Gamma(-\beta)\left(\frac{g_0-g}{g_0}\right)^\beta \longrightarrow \tilde{B}(g)$$
$$= A\Gamma(-\beta-b_1+b_0)\left(\frac{g_0-g}{g_0}\right)^{\beta+b_1-b_0}, \qquad (16)$$

$$B(g) = \frac{A}{\Gamma(\alpha+b_0)}g^\alpha \longrightarrow \tilde{B}(g) = \frac{A}{\Gamma(\alpha+b_1)}g^\alpha.$$

As is seen, an increase in $b_0$ weakens the singularities at a finite point, while the character of singularity at infinity remains unchanged. For sufficiently large $b_0$, the contributions from finite points to $U_N$ are suppressed and the corresponding asymptotic behavior is determined by the singularity of $B(g)$ (and, hence, of $W(g)$) at $g \longrightarrow \infty$:

$$U_N = \frac{W_\infty}{\Gamma(2\alpha)\Gamma(b_0+\alpha)}\left(\frac{4}{a}\right)^\alpha N^{2\alpha-1}, \quad N \longrightarrow \infty. \qquad (17)$$

This formula solves the problem: the coefficients $U_N$ are related by a linear transformation (11) to the initial coefficients $W_N$ (see Eq. (8)), while their asymptotic behavior (17) determine the parameters $W_\infty$ and $\alpha$ of asymptotics (7).

Formulas (14) indicate that a contribution to $U_N$ from the singular point $g = \infty$ is monotonic, while the contributions from other points are oscillating. Therefore, increasing $b_0$ leads to a change in the $U_N$ behavior from oscillating to monotonic. This phenomenon was observed in [6] and, albeit not given any satisfactory explanation, regularly employed for improving the divergence of perturbative series.

### *2.2. Modified Conformal Mapping*

A more effective algorithm is provided by using a modified conformal mapping.

According to the Riemann theorem [21], the conformal mapping of a simply connected region into a unity circle is unique to within the so-called normalization, which can be fixed by setting the images of two (internal and boundary) points. Under the convention that the point $g = 0$ is imaged by $u = 0$ and $g = \infty$ by $u = 1$, conformal mapping (10) is the only one that allows the Borel image to be analytically continued to arbitrary complex $g$ values. However, this is not necessary: to perform the integration in (8), the analytical continuation to positive semiaxis is sufficient. Then, any conformal mapping into a region of the type depicted in Fig. 1c is admissible, in which the point $u = 1$ is the boundary point closest to the origin. The second series in expansion (9) is convergent at $u < 1$ and, in particular, in the interval $0 < u < 1$ imaging the positive semiaxis. An advantage of this conformal mapping is that the contributions from singular points $P$, $Q$, $Q'$, $R$, $R'$ … to $U_N$ are exponentially suppressed and the $U_N$ asymptotics for all $b_0$ is determined by a contribution of the singular point at $u = 1$ related to the singularity of $W(g)$ at $g \longrightarrow \infty$.

Let us use an extremal form of such mapping, imaging the plane with cut $(-\infty, -1/a)$ into the plane with cut $(1, \infty)$ (Fig. 1d). This mapping is given by the formula

$$g = \frac{u}{a(1-u)}, \qquad (18)$$

which leads to the following relationship between $U_N$ and $B_N$:

$$U_0 = B_0,$$
$$U_N = \sum_{k=1}^{N} \frac{B_K}{a^K}(-1)^K C_{N-1}^{K-1} \quad (N \geq 1). \qquad (19)$$

The asymptotic behavior of $U_N$ for large $N$ is

$$U_N = U_\infty N^{\alpha-1}, \quad N \longrightarrow \infty, \qquad (20)$$

$$U_\infty = \frac{W_\infty}{a^\alpha \Gamma(\alpha)\Gamma(b_0+\alpha)}. \qquad (21)$$

As a result, we arrive at a simple algorithm: calculate coefficients $B_N$ by formula (8) using preset $W_N$, recalculate $B_N$ to $U_N$ using relationship (19), and take the power limit (20) for large $N$ to determine parameters $W_\infty$ and $\alpha$ for asymptotics (7).

### *2.3. Random Error Growth*

The above algorithms possess an implicit drawback that significantly restricts the accuracy of description. Let us introduce a reduced coefficient function:

$$F_N = \frac{W_N}{W_N^{as}} = \frac{W_N}{ca^N\Gamma(N+b)}$$
$$= 1 + \frac{A_1}{N} + \frac{A_2}{N^2} + \ldots + \frac{A_K}{N^K} + \ldots, \qquad (22)$$

which varies within finite limits and admits a regular expansion in the powers of $1/N$. The latter can be checked by calculating sequential corrections to the Lipatov asymptotics [19]. In practice, $F_N$ is set with a certain accuracy $\delta_N$ (calculation or round-off error), which leads to a random error in $U_N$. The error dispersion for the algorithm considered in Section 2.2 is as follows:

$$(\delta U_N)^2 = \sum_{K=1}^{N}\left[c\delta_K \frac{\Gamma(K+b)}{\Gamma(K+b_0)} C_{N-1}^{K-1}\right]^2. \qquad (23)$$

For the round-off errors, the value of $\delta_K = \delta$ is independent of $K$. A sum calculated by the steepest descent method for large $N$,

$$\delta U_N \sim 2^N \delta, \qquad (24)$$

demonstrates a catastrophic growth of the error. Calculation with a double computer accuracy yields $\delta \sim 10^{-14}$, so that $\delta U_N$ is on the order of unity for $N \approx 45$.[2] This restricts the accuracy of determining the parameters of asymptotics (7) to approximately 1%. According to expression (23), an increase in $b_0$ decreases the error so that the permissible $N$ level grows. However, large $b_0$ values delay the process of attaining the asymptote (20), so that no advantages are eventually gained.

For the algorithm considered in Section 2.1, the error grows at a still higher rate,

$$\delta U_N \sim (\sqrt{2}+1)^{2N}\delta \sim 5.8^N \delta, \qquad (25)$$

and the requirement of using sufficiently large $b_0$ significantly restricts the possibility of optimization (see Section 3). Nevertheless, this algorithm may still be useful to increase the accuracy of calculations in the region of small $g$ (Section 7). Below we dwell on the algorithm of Section 2.2 based on a modified conformal mapping, which offers indisputable advantages in the region of strong coupling.

[2] This error growth is observed in fact in the form of rapidly increasing irregular oscillations.

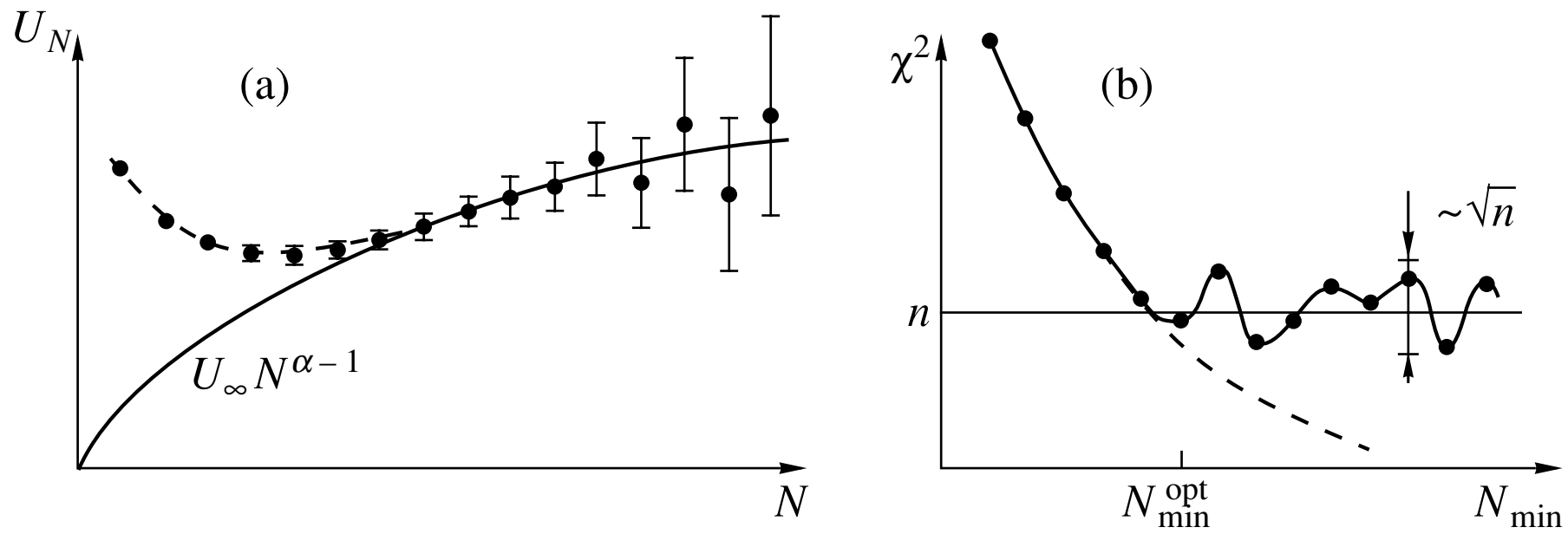

**Fig. 2.** The $U_N$ treatment according to the power law: (a) a typical situation whereby large $N$ correspond to a large statistical error and small $N$, to a large systematic error; (b) the plot of $\chi^2$ versus $N_{\min}$ at a constant number of points $n$.

The above considerations indicate that the computer round-off errors restrict the accuracy of the algorithm to ~1% even for test examples where the $W_N$ values are precisely known. In real cases, the accuracy of $W_N$ calculations is much worse and the situation might appear as hopeless. However, this is not so in fact because we mostly deal with interpolation errors, the influence of which has a quite different character. The linear relationship (19) known in mathematics as the Hausdorff transformation [20] possesses a remarkable property

$$\sum_{K=1}^{N} K^m (-1)^K C_{N-1}^{K-1} = 0, \quad m = 0, 1, \ldots, N-2, \quad (26)$$

that makes smooth errors (well approximated by polynomials) insignificant even despite their large magnitude.[3] Of course, limitations related to the computer round-off error are still valid, but a 1% accuracy is quite sufficient for real problems and this level can hardly be improved for the level of information accessible at present.

Strictly speaking, the problem of round-off errors is purely technical and can be solved by means of special precise arithmetic programs which allow the calculations to be performed with arbitrary number of significant digits [22], however, the accuracy of $\alpha$ and $W_\infty$ restoration logarithmically depends on the computation accuracy. Algorithms that are more perfect in this respect do exist, but their consideration falls outside the scope of this paper; such methods, albeit providing for a high accuracy in the test examples, are insufficiently robust and work unsatisfactorily under conditions of restricted information. The algorithm under consideration is quite stable and, in the author's opinion, ideally suited to obtaining a reliable zero-order approximation.[4]

Treating $U_N$ by the power law can involve a standard procedure of minimization of $\chi^2$ [22]:

$$\chi^2 = \sum_i \left( \frac{y_i - y(x_i)}{\sigma_i} \right)^2, \quad (27)$$

where $y_i$ are the values set at the points $x_i$ with a statistical error $\sigma_i$ and fitted to the theoretical function $y(x)$. In this process, it is important to select properly the interval $N_{\min} \le N \le N_{\max}$ for the $U_N$ treatment. Indeed, large $N$ values lead to large statistical errors determined by formula (23), while small $N$ values increase the role of a systematic error related to the fact that $U_N$ still did not attain asymptotics (20) (Fig. 2a). The upper limit $N_{\max}$ can be chosen using the condition $\delta U_N \sim U_N$, since the points with greater $N$ provide no additional information; this choice is not very critical since the procedure of $\chi^2$ minimization automatically discriminates the points with large statistical errors, which are used in averaging with a weight of $1/\sigma_i^2$. The lower limit $N_{\min}$ has to be selected taking into account the $\chi^2$ value, which reaches an extremely high level for small $N_{\min}$ but attains a "normal" level of $n \pm \mathrm{const}\sqrt{n}$ ($n$ is the number of points) with increasing $N_{\min}$ (Fig. 2b). The optimum value of $N_{\min}$ corresponds to the left end of the "plateau," where a systematic error becomes smaller than the statistical error and the available information is most completely employed.

In fact, the conditions for a strict statistical treatment of $\chi^2$ were not fulfilled because the errors $\delta U_N$ for various $N$ were not independent (see Eq. (23)). This was manifested by the fact that $\chi^2$ values decreased below the "normal" level (dashed curve in Fig. 2b), while the statistical uncertainty of $\alpha$ and $W_\infty$ became very small and did not reflect real errors even in the order of magnitude. For this reason, we considered the choice of $N_{min}$ as satisfactory when the $\chi^2$ values were on the correct order of magnitude ($\sim n$); small changes in $N_{min}$ did not significantly influence the results.

[3] This implies that, in the case when many $W_N$ values are known with low precision, the data should be used upon approximation by a smooth function rather than directly.

[4] This situation is well known in computational mathematics [22]. All algorithms can be roughly divided into two groups: those in the first group possess moderate accuracy and convergence rate but are highly reliable (an example is offered by seeking for a root of equation through segment halving); algorithms of the second group show high accuracy and ensure rapid convergence but pose stringent requirements with respect to the function smoothness (e.g., in seeking a root with the forecast for several derivatives).

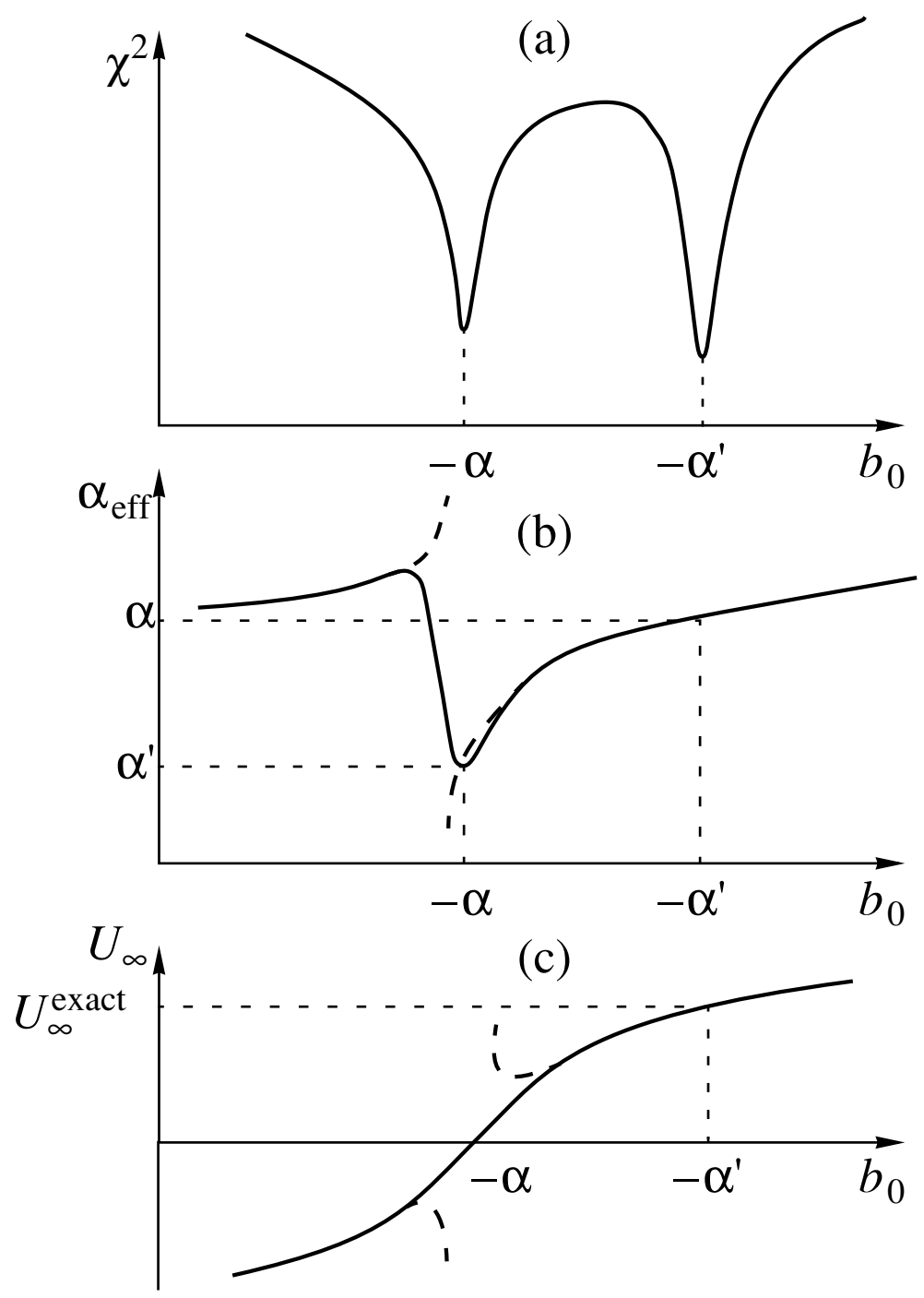

**Fig. 3.** Theoretical plots of (a) $\chi^2$, (b) $\alpha_{eff}$, and (c) $U_\infty$ versus $b_0$ constructed with neglect of the correction terms indicated by dots in expression (29).

## 3. DEPENDENCE ON THE PARAMETER $b_0$ AND ACCURACY ESTIMATION

Direct application of the algorithm described in Section 2.2 is insufficiently effective since the results depend on the arbitrary parameter $b_0$, which implies that an additional investigation is necessary to select the optimum value.

It is naturally expected that corrections to asymptote (7) have the form of a regular expansion with respect to $1/g$. However, even the simplest examples show that, in the general case, this assumption is not valid: in the zero-dimensional case, the corrections follow the powers of $g^{-1/2}$; for an anharmonic oscillator, the corrections follow the powers of $g^{-2/3}$ (see Section 4). For this reason, we admit the power corrections in the general form:

$$W(g) = W_\infty g^\alpha + W'_\infty g^{\alpha'} + \dots . \quad (28)$$

Accordingly, the asymptotic behavior of $U_N$ written by analogy with (20) and (21) is described as

$$U_N = \frac{W_\infty}{a^\alpha \Gamma(\alpha)\Gamma(b_0+\alpha)} N^{\alpha-1} + \frac{W'_\infty}{a^{\alpha'}\Gamma(\alpha')\Gamma(b_0+\alpha')} N^{\alpha'-1} + \dots . \quad (29)$$

First, let us neglect the correction terms indicated by dots in expansion (29). A formal treatment of this expression according to the power law (20) yields quite satisfactory results because the truncated function (29) in the double logarithmic scale varies smoothly and is well approximated by a straight line. However, this approximation only leads to certain effective values of $\alpha$ and $U_\infty$.

Note that, because of the poles of the gamma function, the first and second terms in (29) become zero for $b_0 = -\alpha$ and $b_0 = -\alpha'$, respectively. These $b_0$ values correspond to the purely power laws, $U_N \propto N^{\alpha'-1}$ and $U_N \propto N^{\alpha-1}$, which results in increasing quality of the approximation and a sharply decreasing $\chi^2$ value. Within a fixed working interval $N_{min} \le N \le N_{max}$, the pattern is as follows (Fig. 3): the $\chi^2$ versus $b_0$ curve exhibits sharp minima at $b_0 = -\alpha'$ and $b_0 = -\alpha$; the effective index $\alpha_{eff}$ drops down to $\alpha'$ in the vicinity of $b_0 = -\alpha$ and is close to $\alpha$ outside this region (being exactly equal to $\alpha$ at $b_0 = -\alpha'$); the effective parameter $U_\infty$ corresponds to exact $W_\infty$ at $b_0 = -\alpha'$ and crosses the zero level in the vicinity of the point $b_0 = -\alpha$. The slope of a linear portion of the curve near this root is

$$U_\infty \approx \frac{W_\infty}{a^\alpha \Gamma(\alpha)}(b_0+\alpha), \quad (30)$$

which provides for an $W_\infty$ estimate not too sensitive with respect to $\alpha$ errors. The rejected terms in (29) may only slightly perturb this pattern.

The pattern outlined above was actually observed, but the behavior of $\alpha_{eff}$ and $U_\infty$ in the vicinity of $b_0 = -\alpha$ is usually discontinuous (as indicated by dashed branches in the curves of Fig. 3). However, this circumstance is not physically significant and only reflects features of the mathematical procedure involving taking logarithm of the $U_N$ modulus,

$$\ln|U_N| = \ln|U_\infty| + (\alpha-1)\ln N, \quad (31)$$

followed by using a linear fitting algorithm [22]. The sign of $U_\infty$ is determined by calculating $\chi^2$ for $U_\infty = |U_\infty|$ and $-|U_\infty|$ and selecting a variant with the minimum value. This procedure leads to rather senseless results in the case of $U_N$ changing sign, but this is only possible in a small vicinity of the point $b_0 = -\alpha$, while the sign of $U_N$ outside this narrow interval is determined by the sign of the first term in the right-hand part of Eq. (29).

Smoothness of the $U_\infty(b_0)$ function is restored when the treatment according to power law (20) is performed by varying only $U_\infty$ at a fixed (approximate) $\alpha$ value. Small variations of $\alpha$ virtually do not affect the position of the root of $U_\infty(b_0)$, while significantly influencing the $W_\infty$ value determined from the slope of the linear relationship (30). The above considerations suggest four different methods for estimating the $\alpha$ index, based on (i) the $\alpha_{\text{eff}}$ value at the first minimum of $\chi^2$ (counting from large $b_0$), (ii) the position of the second $\chi^2$ minimum, (iii) the change in the sign of $U_\infty$ upon the logarithmic treatment, and (iv) the change in the sign of $U_\infty$ upon treatment at a fixed $\alpha$ value (taken equal to a preliminary estimate).

The first two estimates ensure, in the general case, a higher precision, since their uncertainty is determined by the ratio of rejected terms in the right-hand part of expansion (29) to the characteristic value of the first term outside the narrow vicinity of $b_0 \approx -\alpha$. The accuracy of the last two estimates is determined by the ratio of the second term to the first term. When the rejected terms in (29) are comparable with the second term (this condition can be monitored by reproducibility of the $\alpha'$ value), all four methods are on the same footing. In practice, it is always important to monitor the change in the sign of $U_\infty$ because this point reliably indicates the minimum in $\chi^2$ corresponding to $b_0 = -\alpha$ (the numbering of minima may change because of their disappearance, appearance of spurious minima, etc. (see below).

There are three possible estimates of $W_\infty$, which use either (i) the $U_\infty$ value at the first minimum of $\chi^2$ or (ii, iii) the slope of a linear portion of the $U_\infty(b_0)$ curve in the vicinity of the root for the treatment at a fixed $\alpha$ (variation of the latter parameter within the interval of $\alpha$ uncertainty obtained by the four methods indicated above provides the upper and lower estimates for $W_\infty$, respectively).

As can be readily shown, a difference between various estimates of $\alpha$ and $W_\infty$ is on the same order of magnitude as the deviation of each estimate from the exact value. This correlation can be used for estimating errors. The availability of several estimates is of great significance: while any two estimated values can accidentally be close to each other (leading to understated value of the predicted error), the accidental proximity of three or four estimates is hardly probable.

## 4. TEST EXAMPLES

The operation of the proposed algorithm can be illustrated by application to several test systems.

### *4.1. Zero-Dimensional Case*

The first example is offered by the integral

$$W(g) = \int_0^\infty d\varphi \varphi^{n-1} \exp(-\varphi^2 - g\varphi^4), \tag{32}$$

which can be considered as a zero-dimensional limit of the functional integral in the $n$-component $\varphi^4$ theory. Here, it is easy to calculate the expansion coefficients

$$W_N = ca^N \frac{\Gamma\left(N + \frac{n+2}{4}\right)\Gamma\left(N + \frac{n}{4}\right)}{\Gamma(N+1)} \tag{33}$$

and their behavior for large $N$:

$$W_N = ca^N \Gamma(N+b)\left\{1 + \frac{A_1}{N} + \dots\right\}, \tag{34}$$

where

$$a = 4, \quad b = \frac{n-1}{2},$$
$$c = \frac{2^{n/2}}{4\sqrt{\pi}}, \quad A_1 = \frac{(n-2)(4-n)}{16}. \tag{35}$$

Asymptotic behavior of the integral at $g \longrightarrow \infty$ is described by the following relationships:

$$W(g) = W_\infty g^\alpha, \quad \alpha = -n/4,$$
$$W_\infty = \Gamma(n/4)/4, \tag{36}$$

with the corrections having the form of a series in powers of $g^{-1/2}$. In the test, the required number of coefficients $W_N$ was set with a double computer accuracy ($\delta \sim 10^{-14}$), after which the $\alpha$ and $W_\infty$ values were restored assuming their Lipatov asymptotics to be known.

**(i) *n* = 1.** Figure 4 shows the $\tilde{U}_N$ against $N$ curves calculated for various values of the parameter $b_0$ (points) and the results of treatment according to the power law (solid curves). For better illustration, the data are presented in the form of coefficients,

$$\tilde{U}_N = U_N \Gamma(b_0 + N_0), \tag{37}$$

normalized so as to tend to a finite limit for $b_0 \longrightarrow \infty$; $N_0$ is the lower limit of summation in relationship (19), which can differ from unity when several first terms of the series (1) are zero. As is seen, all curves in fact exhibit a power asymptotic behavior for large $N$. Attaining the asymptote is delayed for $b_0 \gg 1$ and $b_0 \longrightarrow -N_0$, because of the existence of the corresponding large parameters in relationship (19). In contrast, the power law holds even for small $N$ for $b_0 = 0.82$ corresponding to the first minimum of $\chi^2$.

Figure 5 shows the plots of $\chi^2$, $\alpha_{\text{eff}}$, and $\tilde{U}_\infty = U_\infty \Gamma(b_0 + N_0)$ versus $b_0$ calculated in the interval $24 \le N \le 50$. For the first minimum of $\chi^2$ corresponding to $b_0 = 0.82$, estimates obtained according to Section 3 are as follows:

$$\alpha = -0.247, \quad W_\infty = 0.892, \quad \alpha' = -0.82. \tag{38}$$

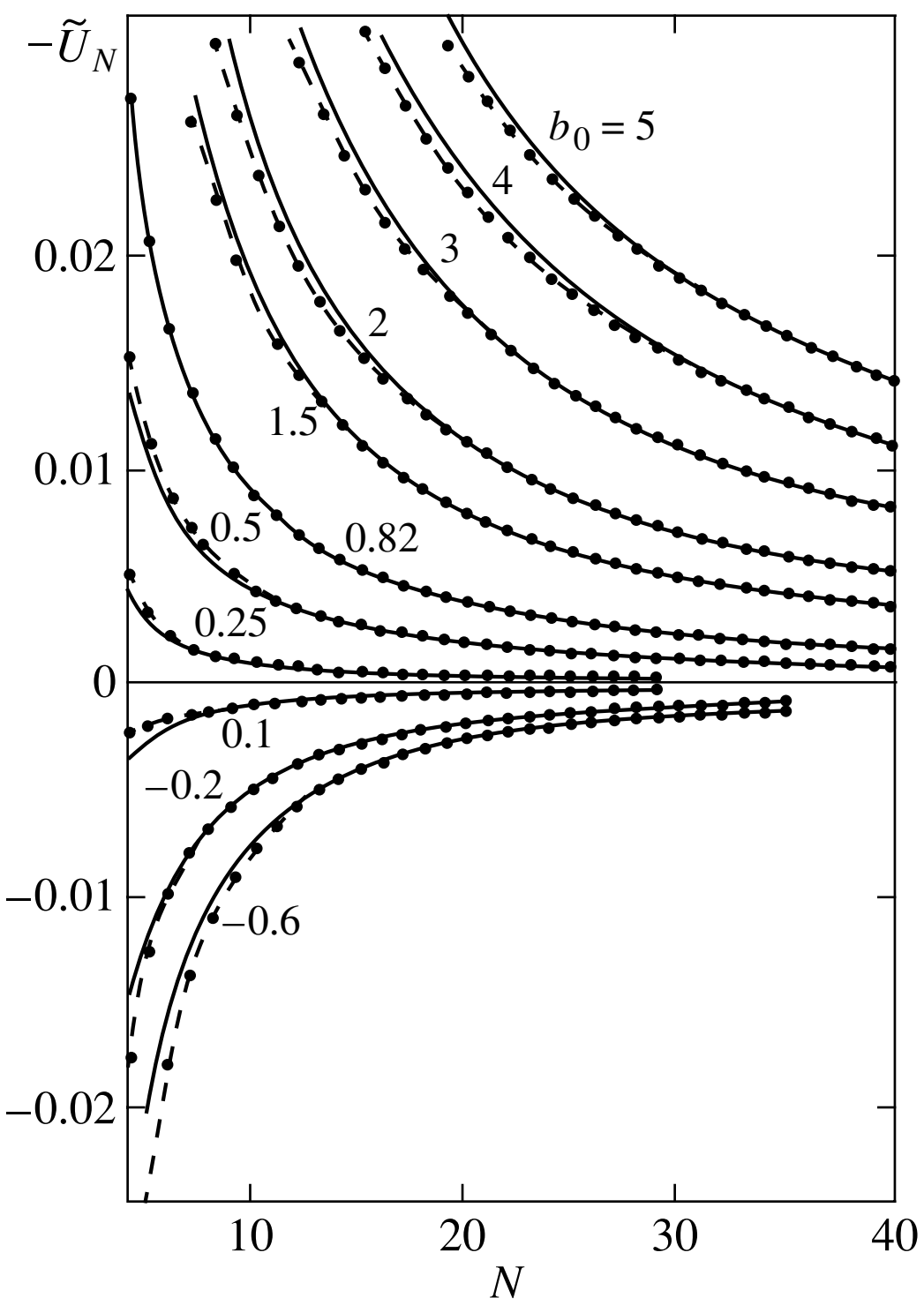

**Fig. 4.** The plots of $\tilde{U}_N = U_N\Gamma(b_0 + 1)$ versus $N$ at fixed $b_0$ (points and dashed curves) for integral (32) with n=1. Solid curves show the results of treatment according to the power law.

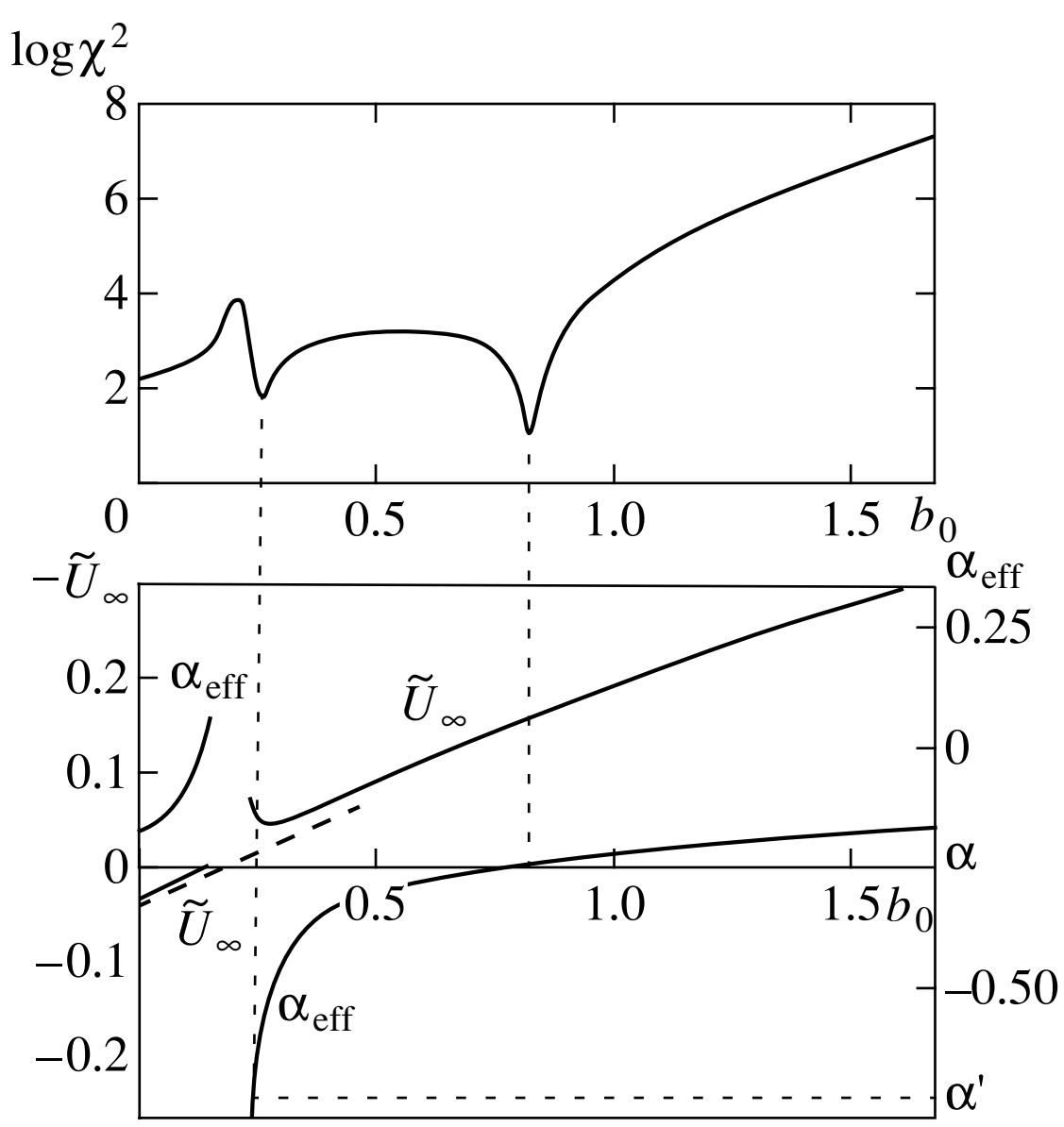

**Fig. 5.** The plots of $\chi^2$, $\alpha_{eff}$, and $\tilde{U}_\infty = U_\infty\Gamma(b_0 + 1)$ versus $b_0$ for integral (32) with $n = 1$ in the averaging interval of $24 \le N \le 50$. Dashed line shows a portion of the $U_\infty(b_0)$ curve in the vicinity of the root, obtained by the treatment at a constant index α = –0.25.

The second minimum of $\chi^2$ taking place at $b_0 = 0.26$ yields

$$\alpha = -0.26, \quad \alpha' = -0.67. \tag{39}$$

The $U_\infty$ value changes sign at $b_0 = 0.210$ and 0.215 for the treatment with taking a logarithm and at a fixed index, which yields the estimates α = –0.210 and –0.215, respectively. The slope of a linear portion in the $U_\infty(b_0)$ curve in the vicinity of the root (dashed line in Fig. 5 constructed upon treatment at a fixed index) yields the $W_\infty$ values depending on the preselected α value: for α = –(0.21–0.26), the estimates range within $W_\infty$ = 0.883–0.933. Summarizing all these estimates, we obtain the set of estimates

$$\alpha = -0.235 \pm 0.025, \quad W_\infty = 0.908 \pm 0.025, \quad \alpha' = 0.75 \pm 0.08, \tag{40}$$

which are consistent with the exact values

$$\alpha = -0.25, \quad W_\infty = 0.9064, \quad \alpha' = -0.75. \tag{41}$$

Since the α' values in (38) and (39) agree satisfactorily, we may conclude that the rejected terms in expansion (29) are small as compared to the second term. Therefore, the best estimates for α are provided (see Section 3) by relationships (38) and (39). Restricting to these estimates, we obtain

$$\alpha = -0.253 \pm 0.007, \quad W_\infty = 0.887 \pm 0.005 \tag{42}$$

instead of set (40). Here, the accuracy of determining α really increased, but the error of $W_\infty$ is somewhat underestimated.

The shape of the $\chi^2$ curves is highly sensitive to selection of the lower boundary of the working interval $N_{min} \le N \le N_{max}$. As the $N_{min}$ value decreases, the $\chi^2$ minima tend to smear, while an increase in $N_{min}$ leads to flattening of the curves and the appearance of small-scale fluctuations hindering identification of the minima. In attempts at obtaining the clearest minima corresponding to $\chi^2$ values of the correct order in magnitude, the choice was usually made between two–three $N_{min}$ values.[5] A change in the working interval most significantly affects the estimates (39), with the α and α' variations approximately corresponding to a difference between (38) and (39).

**(ii) *n* = 2.** The $\chi^2$ plots in Fig. 6 exhibit sharp minima at $b_0 = 1.26$ and 0.50. The first $\chi^2$ minimum yields

$$\alpha = -0.4996, \quad W_\infty = 0.442, \quad \alpha' = -1.26, \tag{43}$$

while the other three methods give α = –0.5000 accurate to within the last digit. An estimate for α' obtained using the second $\chi^2$ minimum amounts to about 20, which is inconsistent with (43). Therefore, the rejected terms in (29) are comparable with the second, so that all four possible estimates are on the same footing. Treatment of a linear portion of the $U_\infty(b_0)$ curve near the root yields $W_\infty = 0.460$. As a result, we obtain

$$\alpha = -0.5000 \pm 0.0004, \quad W_\infty = 0.451 \pm 0.009, \tag{44}$$

in good agreement with the exact values

$$\alpha = -0.50, \quad W_\infty = 0.4431. \tag{45}$$

[5] It should be noted that, in displaying the results of calculations with fixed decimal point, the $\chi^2$ minima are well distinguished by the configuration of digits even in the course of a rapid on-screen computer survey.

**(iii) *n* = 3.** Here, the $\chi^2(b_0)$ plots exhibit minima at $b_0 = 1.07$ and 0.77, which yield

$$\alpha = -0.704, \quad W_\infty = 0.192, \quad \alpha' = 1.07 \tag{46}$$

and

$$\alpha = -0.77, \quad \alpha' = -1.42, \tag{47}$$

respectively. Estimates obtained using $U_\infty$ changing sign are $\alpha = -0.86$ for the treatment with taking a logarithm and $\alpha = -0.84$ for the treatment at a fixed index. Determining $W_\infty$ from the slope of a linear portion in the $U_\infty(b_0)$ curve in the vicinity of the root yields 0.311, 0.420, and 0.751 for $\alpha = -0.704$, $-0.77$, and $-0.86$, respectively. Since the two values of $\alpha'$ reasonably agree with each other, the estimates (46) and (47) for $\alpha$ must be more precise. Taking only these estimates into account, we obtain

$$\alpha = -0.737 \pm 0.033, \quad W_\infty = 0.306 \pm 0.114, \quad \alpha' = -1.25 \pm 0.18, \tag{48}$$

in good agreement with the exact values

$$\alpha = -0.75, \quad W_\infty = 0.3063, \quad \alpha' = -1.25. \tag{49}$$

An allowance for all four estimates of $\alpha$ yields

$$\alpha = -0.78 \pm 0.08, \quad W_\infty = 0.47 \pm 0.28 \tag{50}$$

with markedly greater errors.

In this case, we may also point out difficulties arising due to an additional "spurious" minimum appearing at $b_0 = 1.90$. However, this minimum can be excluded from consideration upon identifying the minimum at $b_0 = 0.77$ as corresponding to $b_0 = -\alpha$ (by $U_\infty$ changing sign) and the minimum at $b_0 = 1.07$ as corresponding to $b_0 = -\alpha'$ (by the consistent $\alpha'$ values). In the general case, the process of identifying useful minima resembles the situation in spectroscopy under high noise conditions: selecting informative signals requires certain skill.

**(iv) *n* = 4.** In this case, application of the algorithm encounters the "hidden rock" of this method. Based on the usual estimates, we obtain a quite precise result:

$$\alpha = -1.500 \pm 0.004, \quad W_\infty = -0.222 \pm 0.005. \tag{51}$$

However, these values do not agree with (36). The discrepancy is caused by the fact that the main contribution to the $U_N$ asymptotics vanish because the gamma function exhibits a pole at the exact value of the index $\alpha = -1$ (see Eq. (29)), so that the next term of the expansion becomes significant with the parameters

$$\alpha' = -1.50, \quad W'_\infty = -\sqrt{\pi}/8 = -0.2216. \tag{52}$$

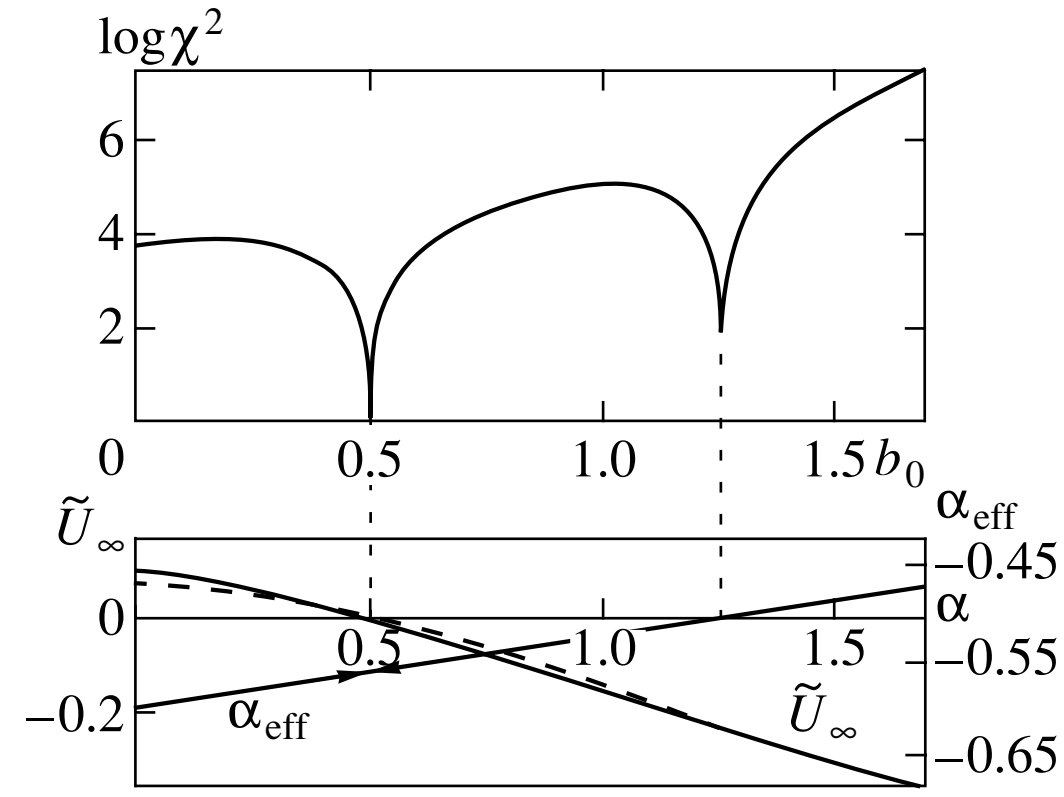

**Fig. 6.** The plots of $\chi^2$, $\alpha_{\rm eff}$, and $\tilde{U}_\infty = U_\infty\Gamma(b_0 + 1)$ versus $b_0$ for integral (32) calculated with $n = 2$ in the averaging interval of $20 \le N \le 50$. Dashed line shows a portion of the $U_\infty(b_0)$ curve in the vicinity of the root, obtained by the treatment at a constant index $\alpha = -0.5$. The $\alpha_{\rm eff}$ for $b_0 = 0.5$ falls far outside the diagram boundaries.

Therefore, the proposed algorithm is incapable of restoring correct asymptotics described by Eq. (7) in the case of nonpositive integer $\alpha$ values. In order to avoid these problems, the algorithm has to be supplemented by the following rule: if the treatment yields a negative $\alpha$ value, the result must be checked by taking a negative or fractional power of series (1) and summing the reexpanded series.

### *4.2. Anharmonic Oscillator*

The second example is offered by the problem of determining the ground sate $E_0(g)$ of an anharmonic oscillator described by the Schrödinger equation

$$\left\{-\frac{d^2}{dx^2} + \frac{x^2}{4} + \frac{gx^4}{4}\right\}\psi(x) = E\psi(x). \tag{53}$$

This problem can be reduced to a one-dimensional $\varphi^4$ theory. Consider $E_0(g)$ as the $W(g)$ function with the initial terms of the perturbative series having the following form:

$$W(g) = \frac{1}{2} + \frac{3}{4}g - \frac{21}{8}g^2 + \frac{333}{16}g^3 - \frac{30885}{128}g^4 + \ldots. \tag{54}$$

Bender and Wu [23] calculated the first 75 coefficients $W_N$ up to the 12th decimal digit and obtained an expres-

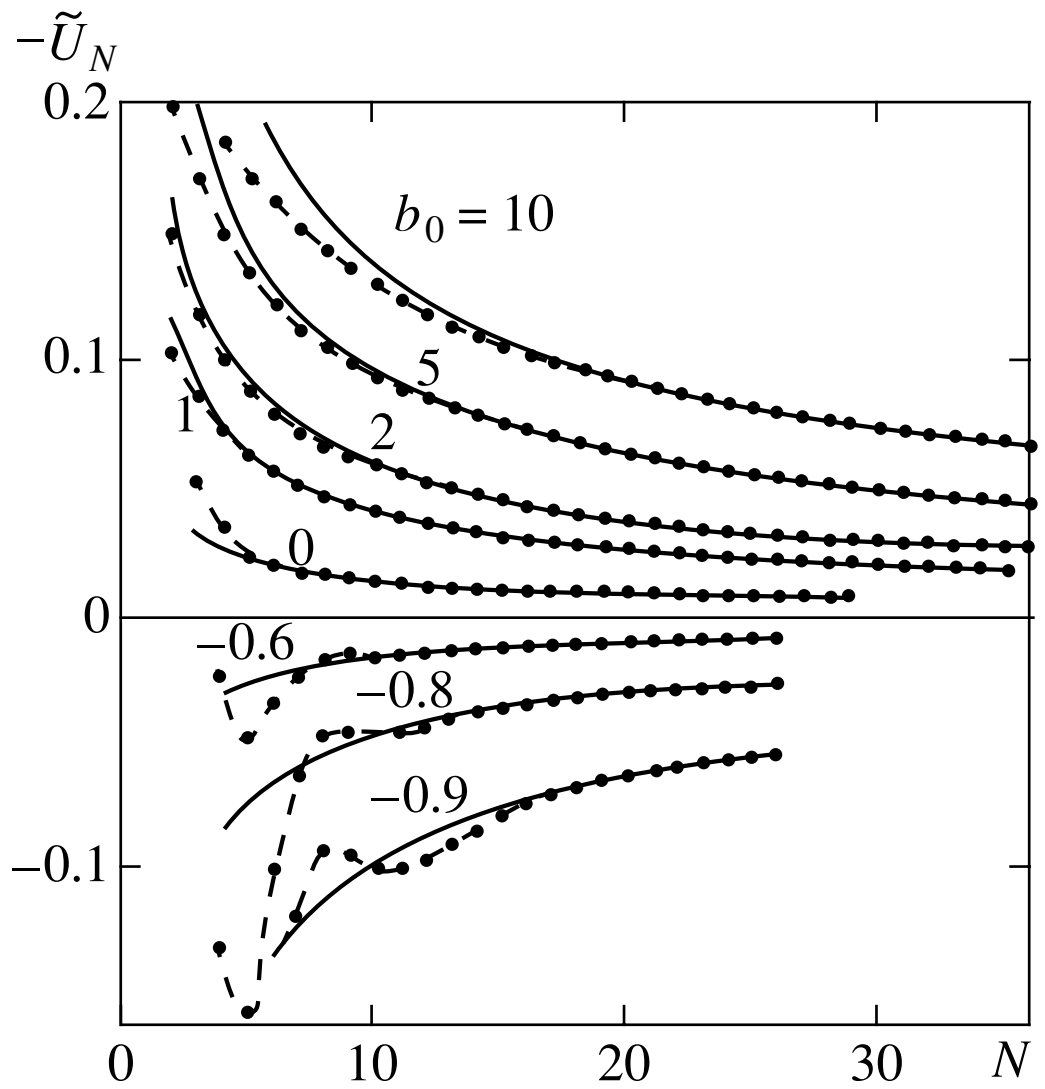

**Fig. 7.** The plots of $\tilde{U}_N = U_N\Gamma(b_0 + 1)$ versus $N$ for an anharmonic oscillator. The notations are the same as in Fig. 4.

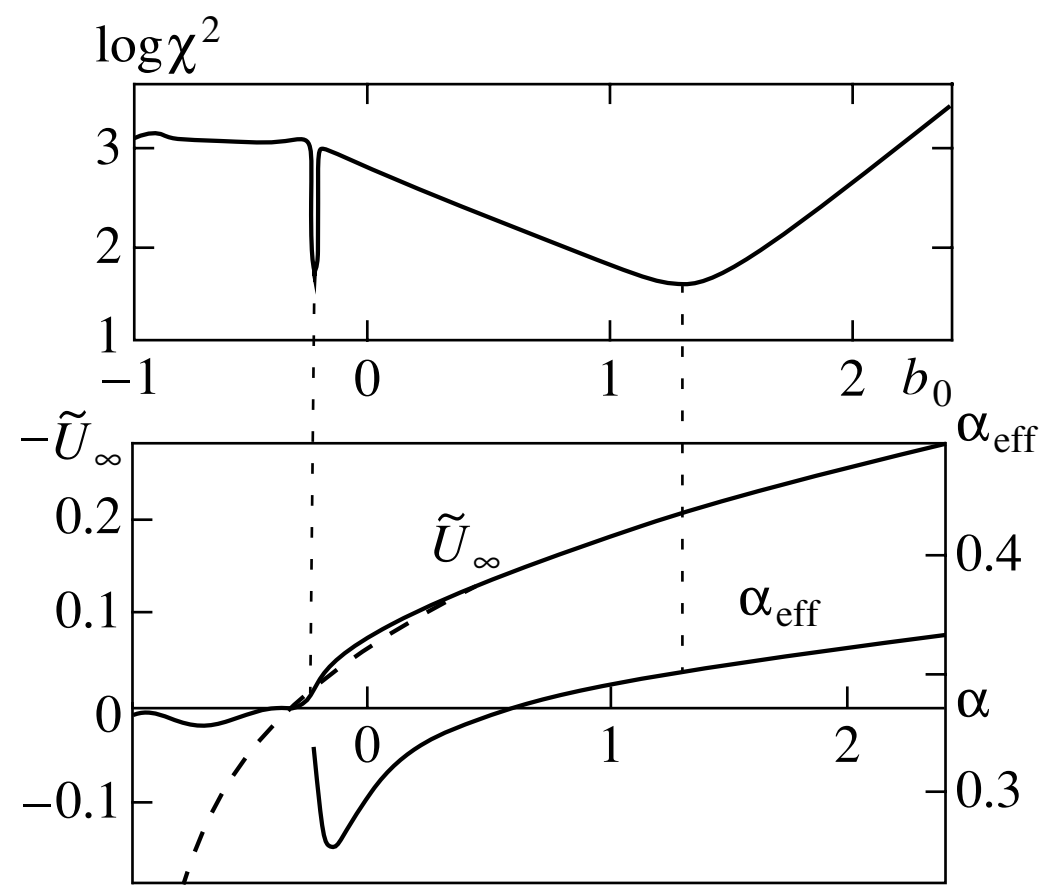

**Fig. 8.** The plots of $\chi^2$, $\alpha_{eff}$, and $\tilde{U}_\infty = U_\infty\Gamma(b_0 + 1)$ versus $b_0$ for an anharmonic oscillator in the averaging interval of $24 \le N \le 45$. Dashed line shows the result of treatment at a constant index $\alpha = 0.34$.

sion describing behavior of the expansion coefficients with large $N$:

$$W_N = -\frac{\sqrt{6}}{\pi^{3/2}}3^N\Gamma\left(N+\frac{1}{2}\right)\left\{1-\frac{95/72}{N}+\dots\right\}. \qquad (55)$$

The asymptotics of $E_0(g)$ for $g \longrightarrow \infty$ is revealed by substituting $E_0(g) = \lambda_0 g^{1/3}$and $x \longrightarrow xg^{-1/6}$, after which Eq. (53) transforms into

$$\left\{-\frac{d^2}{dx^2}+\frac{x^2}{4}+\frac{x^2}{4g^{2/3}}\right\}\psi(x) = \lambda_0\psi(x). \qquad (56)$$

For $g \longrightarrow \infty$, the last term in braces is insignificant and $\lambda_0$ tends to a constant value of 0.6679863 that can be determined by the variational method [24]. Thus, the $W(g)$ asymptotics is described by power series (7) with the parameters

$$\alpha = 1/3, \quad W_\infty = 0.668, \qquad (57)$$

and the corrections having the form of a series in powers of $g^{-2/3}$.

Figure 7 presents the plots of $\tilde{U}_N$ against $N$ and the results of their treatment according to the power law. Figure 8 shows the plots of $\chi^2$, $\alpha_{eff}$, and $\tilde{U}_\infty$ versus $b_0$. As is seen, $\chi^2$ exhibits minima $b_0 = 1.30$ and –0.34 corresponding to

$$\alpha = 0.349, \quad W_\infty = 0.602, \quad \alpha' = -1.80 \qquad (58)$$

and

$$\alpha = 0.34, \quad \alpha' \approx 20, \qquad (59)$$

respectively. Estimates obtained using $U_\infty$ changing sign are $\alpha = 0.285$ for the treatment with taking logarithm and $\alpha = 0.337$ for the treatment at a fixed index. Determining $W_\infty$ from the slope of a linear portion in the $U_\infty(b_0)$ curve in the vicinity of the root yields values in the interval from 0.616 to 0.883. The two values of α' having nothing in common indicates that all α estimates are on the same footing. As a result, we obtain

$$\alpha = 0.317 \pm 0.032, \quad W_\infty = 0.74 \pm 0.14, \qquad (60)$$

in good agreement with the exact values (57).

The above examples show that the accuracy of restoring the $W(g)$ asymptotics, while depending significantly on the particular problem, is generally correlated with the character of corrections to the $U_N$ asymptotics described by relationship (20). An average accuracy on the order of $10^{-2}$ is attained in the zero-dimensional case with odd $n$, where the corrections to (20) have the form of power series in $N^{-1/2}$. For even $n$, every other correction vanishes due to the poles of the gamma function to leave a regular expansion in $1/N$, which markedly increases the resulting accuracy. A relatively low accuracy in the case of an anharmonic oscillator is related to the fact that corrections have the form of series in powers of $N^{-1/3}$.[6] It is important to note, however, that the algorithm automatically yields an estimate of the error. The estimate is rather reliable when all four possible methods for evaluating α are employed.

[6] The first term in (28) gives, in addition to the main contribution to $U_N$ proportional to $N^{\alpha-1}$, the regular corrections $N^{\alpha-2}$, $N^{\alpha-3}$, …; the second term contributes by $N^{\alpha'-1}$, $N^{\alpha'-2}$, …, etc. As a result, the expansion in $g^{-2/3}$ converts into the expansion in $N^{-1/3}$.

## 5. ALGORITHM OPERATING WITH INTERPOLATED COEFFICIENT FUNCTION

The importance of interpolation was strongly underestimated, although this method can obviously provide for an increase in the accuracy of calculations. In most investigations in the field under consideration, the algorithms were formulated so as to avoid mentioning the coefficients $W_N$ at intermediate $N$ values. This approach is conceptually incorrect since, using a finite number of the initial coefficients and their asymptotics, it is possible to construct a function with preset behavior in infinity.[7] A reasonable problem formulation corresponds to approximately setting all $W_N$, after which $W(g)$ can be reconstructed with certain precision.

Thus, a necessary stage in solving the problem consists in interpolating the coefficient function, which naturally implies that this function is analytical (see Section 8.2). The interpolation stage allows the parameter $c$ in the Lipatov asymptotics (essentially not used in the standard conform-Borel procedure [6]) to be effectively employed. In addition, it is possible to take into account smoothness of the reduced coefficient function, its regularity with respect to $1/N$, and (eventually) the information concerning asymptotics of the $A_K$ coefficients in expansion (22) [25].

In Section 2.3, some qualitative considerations were presented suggesting that the influence of the interpolation errors is not as significant as that of the round-off errors. Unfortunately, no particular estimates illustrating this were obtained. Validity of this statement will be experimentally demonstrated for the zero-dimensional test example with $n = 1$.

With a view to modeling a situation for the $\varphi^4$ theory, let us assume that several coefficients in the expansion of series (1) are known,

$$W_{L_0},\ W_{L_0+1},\ \ldots,\ W_L, \qquad (61)$$

together with the Lipatov asymptotics (2) and the corresponding first correction in $1/N$. The interpolation is conveniently performed for the reduced coefficient function, retaining a finite number of terms in expansion (22) and selecting coefficients $A_K$ by correspondence to set (61).

Let us consider in detail two examples of the interpolation procedure, which correspond to (i) $L_0 = 1$, $L = 5$ and (ii) $L_0 = 1$, $L = 1$. Owing to a slow character of variation of the coefficient function, the accuracy of interpolation in both cases is very high: $\sim 10^{-9}$ and $\sim 10^{-4}$, respectively. A random error of such amplitude should lead to large fluctuations in $U_N$ for $N \approx 30$ in the former case and $N \approx 13$ in the latter case. Real calculations indicate that no catastrophic consequences take place up to $N = 40$, when the influence of the round-off errors becomes significant. This can be seen in Table 1 presenting the values of some coefficients $U_N$ calculated for $b_0 = 1$ using the exact and interpolated coefficients of $W_N$. An increase in the $b_0$ value improves the accuracy; when $b_0$ decreases, the accuracy drops somewhat, although the resulting deviations would be indistinguishable on the scale of Fig. 4.

**Table 1.** Comparison of $U_N$ values calculated for $b_0 = 1$ using exact and interpolated coefficients $W_N$

| $N$ | $U_N$ | | |
|---|---|---|---|
| | Exact $W_N$ values | Interpolation with $L_0 = 1$, $L = 5$ | Interpolation with $L_0 = 1$, $L = 1$ |
| 30 | $-2.911 \times 10^{-3}$ | $-2.911 \times 10^{-3}$ | $-2.868 \times 10^{-3}$ |
| 35 | $-2.408 \times 10^{-3}$ | $-2.409 \times 10^{-3}$ | $-2.369 \times 10^{-3}$ |
| 40 | $-2.038 \times 10^{-3}$ | $-2.041 \times 10^{-3}$ | $-2.004 \times 10^{-3}$ |

**Table 2.** The parameters of asymptotics for integral (32) with $n = 1$ calculated using exact and interpolated coefficients $W_N$

| Estimates based on | Exact $W_N$ values | Interpolation with $L_0 = 1$, $L = 5$ | Interpolation with $L_0 = 1$, $L = 1$ |
|---|---|---|---|
| First $\chi^2$ minimum | $\alpha = -0.246$ | $\alpha = -0.245$ | $\alpha = -0.269$ |
| | $\alpha' = -0.827$ | $\alpha' = -0.830$ | $\alpha' = -0.761$ |
| | $W_\infty = 0.893$ | $W_\infty = 0.892$ | $W_\infty = 0.912$ |
| Second $\chi^2$ minimum | $\alpha = -0.249$ | $\alpha = -0.245$ | $\alpha = -0.271$ |
| | $\alpha' = -0.792$ | $\alpha' = -0.849$ | $\alpha' = -0.747$ |
| $U_\infty$ changing sign | $\alpha = -0.210$ | $\alpha = -0.210$ | $\alpha = -0.218$ |
| $U_\infty(b_0)$ linear | $\alpha = -0.215$ | $\alpha = -0.215$ | $\alpha = -0.225$ |
| | $W_\infty = 0.889$ | $W_\infty = 0.887$ | $W_\infty = 0.885$ |

The curve of $\chi^2(b_0)$ is analogous to (albeit not fully coinciding with) that depicted in Fig. 5. Estimates of the asymptotic parameters are listed in Table 2; for better illustration, all values refer to the same working interval of $23 \le N \le 45$ and the value $\alpha = -0.25$ used for the treatment of a linear portion of the $U_\infty(b_0)$ curve. As is seen from these data, changes in $\alpha$ and $W_\infty$ caused by the interpolation fall within the scatter of various estimates and virtually do not influence the accuracy of restoration of asymptotics (7). Therefore, interpolation using a single expansion coefficient $W_1$ allowed the $W(g)$ asymptotics to be restored with an accuracy not worse than that achieved with the exact coefficients $W_N$. Of course, this is by no means a typical situation.

[7] A function of the factorial series possesses the same asymptotics of coefficients (2) but with a different parameter $c$ [17]; the last statement in the text can be readily proved by taking an appropriate linear combination of several functions.

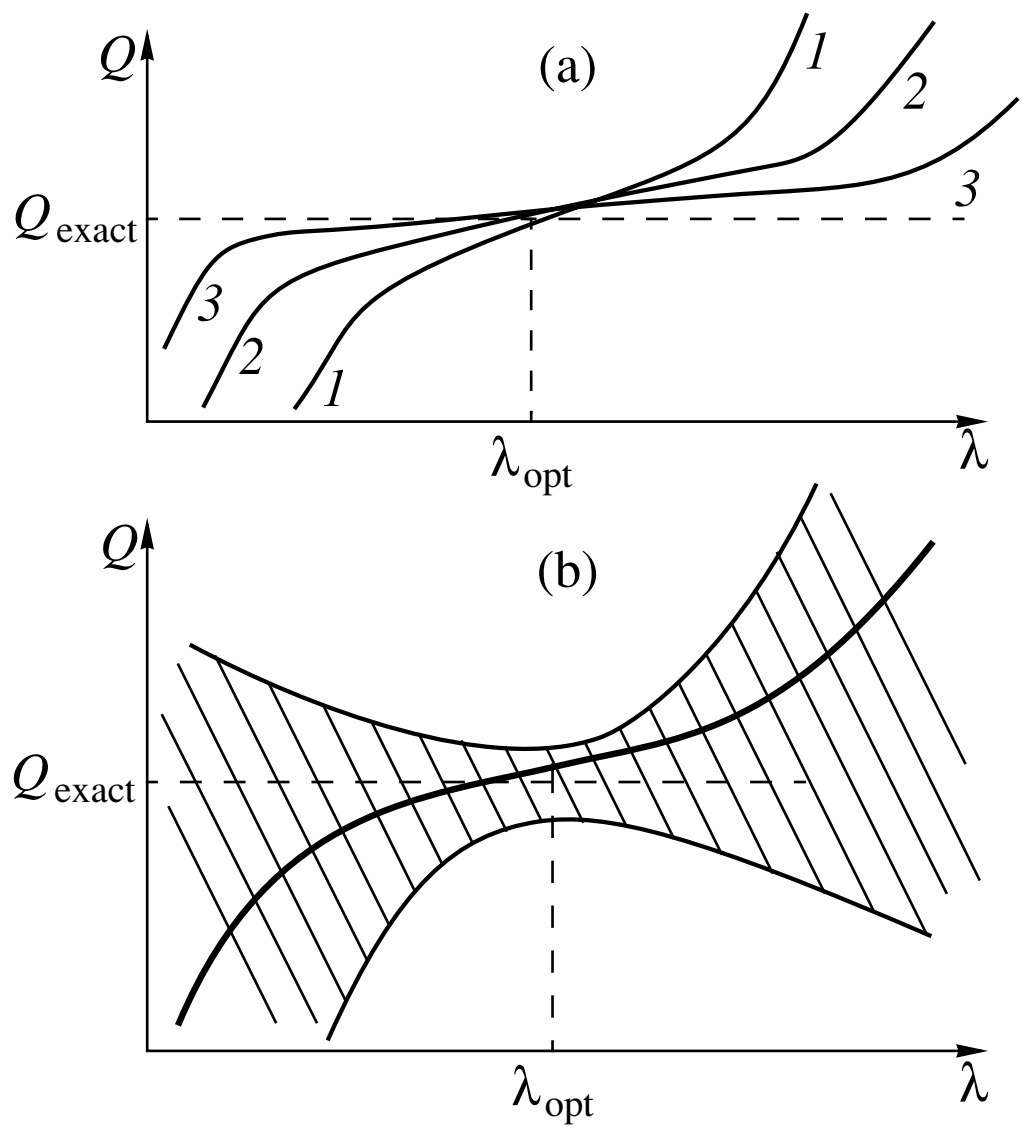

**Fig. 9.** Schematic diagrams illustrating the optimization procedure: (a) theoretically, any quantily $Q$ obtained upon summation of the series must be independent of the optimization parameter $\lambda$; however, such dependence arises under the conditions of restricted information and weakens (on the passage from curve *1* to *2*, *3*, etc.) as the amount of information increases (the optimum value $\lambda = \lambda_{\text{opt}}$ occurs at the center of the plateau); (b) the choice of $\lambda$ affects both the approximate $Q$ value (thick solid curve) and the error of determination (cross-hatched area), so that a correct estimation of this error must provide for the exact value $Q_{\text{exact}}$ being compatible with all data. In the "ideal" situation depicted, optimization with respect to $\lambda$ consists in selecting the result characterized by a minimum error.

## 6. OPTIMIZATION OF THE INTERPOLATION PROCEDURE

Considering an example in the preceding section, we were lucky to see that the most natural method of interpolation may give good results. In the general case, the interpolation procedure requires optimization that will be demonstrated in the case of an anharmonic oscillator. Let us first discuss the general strategy of optimization, which has been significantly modified in comparison to that used in the previous works.

### *6.1. General Strategy of Optimization*

On an abstract level, the optimization consists in introducing a certain variation of the summation procedure characterized by a parameter $\lambda$, the latter value being eventually selected in a "optimum manner." For example, the initial series (1) can be raised to the $\lambda$ power and reexpanded to yield

$$W^{\lambda}(g) = \tilde{W}_0 - \tilde{W}_1 g + \tilde{W}_2 g^2 - \dots + \tilde{c} a^N \Gamma(N+b)(-g)^N + \dots. \tag{62}$$

The properties of this series are analogous to those of the initial one, except for a change in the Lipatov asymptotic parameter $c$ [17]. The new series is summed upon selecting the $\lambda$ value so as to provide for the best convergence of the second series in expansion (9). The optimization procedure is employed, bringing both advantages and troubles, in most investigations in the field under consideration. On the one hand, the principal possibility of improving the convergence is definitely valuable. On the other hand, the results become dependent of an arbitrary parameter $\lambda$ and it is difficult to get rid of the feeling that any result can be obtained.

Theoretically, the use of series (62) is fully equivalent to the study of initial series (1) and the value of any quantity $Q$ obtained upon summation must be independent of the parameter $\lambda$. However, under the conditions of restricted information concerning coefficients $W_N$, the $Q$ value begins to depend on the choice of $\lambda$, this dependence weakening as the amount of information increases. In the general case, no uniform convergence with respect to $\lambda$ takes place and an approximate $Q$ value is close to the exact one only within a certain "plateau" region (Fig. 9a), the deviations rapidly growing outside this region. As the amount of necessary information increases, the plateau expands and flattens (see, e.g., [26]). Apparently, the best convergence takes place at the center of the plateau. However, this point is not always unambiguously selected, since the plateau may be asymmetric or poorly pronounced, the center may shift in the course of convergence, etc. Therefore, selecting the best approximation for $Q$ and estimating the approximation uncertainty are rather subjective procedures.

In the author's opinion, the optimization problem can nevertheless be solved objectively. Indeed, since the choice of $\lambda$ affects both the approximate $Q$ value and the error of determination, a correct estimation of this error must provide for the exact value $Q_{\text{exact}}$ being compatible with the approximate values obtained for any $\lambda$ (Fig. 9b). This criterion eliminates the problem of an apparent dependence of $Q$ on $\lambda$. Once such an "ideal" situation is attained, optimization of the procedure with respect to $\lambda$ reduces to selecting the result characterized by a minimum error.

The optimization procedure is expediently performed in the interpolation stage, since all the final errors arise essentially from the uncertainties in $W_N$. Rewriting expansion (22) in the equivalent form

$$W_N = c a^N N^{\tilde{b}} \Gamma(N + b - \tilde{b}) \times \left\{ 1 + \frac{\tilde{A}_1}{N - \tilde{N}} + \frac{\tilde{A}_2}{(N-\tilde{N})^2} + \dots + \frac{\tilde{A}_K}{(N-\tilde{N})^K} + \dots \right\} \tag{63}$$

and using the interpolation by truncating the series and selecting coefficients $\tilde{A}_K$, we obtain a manifold of realizations of the interpolation procedure characterized by two parameters, $\tilde{b}$ and $\tilde{N}$. An analysis of the test examples shows this parametrization to be sufficiently effective: the accuracy of interpolation achieved for the opti-mum $\tilde{b}$ and $\tilde{N}$ values can be higher by several orders of magnitude as compared to that for a random choice of these parameters. Below, the optimization with respect to $\tilde{b}$ is based on theoretical consideration, while the optimum $\tilde{N}$ value is selected based on the results of numerical calculations.[8]

### 6.2. Optimization with Respect to $\tilde{b}$

Optimization with respect to $\tilde{b}$ is related to the problem of selecting parametrization for the Lipatov asymptotics which can be written in various forms: $ca^N\Gamma(N + b)$, $ca^N N^{b-1} N!$, etc. This problem was actively discussed (see, e.g., [11, 12]), but no satisfactory solutions were proposed.

Note that the values $\tilde{b} = b$ and $\tilde{b} = b - 1$ lead to identical results:

$$N^{\tilde{b}}\Gamma(N + b - \tilde{b}) = \begin{cases} N^b\Gamma(N), & \tilde{b} = b \\ N^{b-1}\Gamma(N+1) = N^b\Gamma(N), & \tilde{b} = b - 1. \end{cases} \quad (64)$$

Therefore, the approximate values of any quantity $Q$ obtained upon summation of the series will coincide for $\tilde{b} = b$ and $b - 1$. As the amount of information concerning the coefficients $W_N$ increases, the $Q(\tilde{b})$ function varies more and more slowly. When the characteristic scale $L$ of this variation increases, the $k$th derivative of the function drops as $1/L^k$. As a result, an extremum at the point $\tilde{b} = b - 1/2$ appears in the general situation, with a plateau between the $Q$ values corresponding to $\tilde{b} = b$ and $b - 1$and the point $\tilde{b} = b - 1/2$ being the natural center of this plateau. The error of restoring $Q$, like any other value, exhibits an extremum (which is naturally expected to be minimum) at $\tilde{b} = b - 1/2$ (see Section 8). Thus, the optimum choice is $\tilde{b} = b - 1/2$; this corresponds to the following parametrization of the Lipatov asymptotics:

$$W_N^{as} = ca^N N^{b-1/2}\Gamma(N + 1/2). \quad (65)$$

The first correction $A_1/N$ to this asymptotics (see expansion (22)) depends on $\tilde{b}$ as

$$A_1 = \bar{A}_1 - (b - 1/2 - \tilde{b})^2/2, \quad (66)$$

where $\bar{A}_1$ is the value of $A_1$ for $\tilde{b} = b - 1/2$. In all known cases, $\bar{A}_1 < 0$ (see [19, 23, 27, 28]) and a minimum correction corresponds to parametrization (65), which favors a good matching between the high-order asymptotics and the low-order behavior. Note that the asymptote according to the Lipatov method [2] is

$$\sqrt{2\pi}c(a/e)^N N^{b-1/2} N^N.$$

The above parametrization (65) corresponds to approximation

$$\sqrt{2\pi}e^{-N}N^N \approx \Gamma(N + 1/2)$$

whose accuracy is 4% even for $N = 1$; so, this parametrisation is close to "natural" one. For an anharmonic oscillator, the optimum parametrization coincides with (55), while in the zero-dimensional case with $n = 1$ it is close to (34) and (35).

### 6.3. Optimization with Respect to $\tilde{N}$

The case of an anharmonic oscillator was studied in detail using the interpolation with $L_0 = 1$, $L = 9$ (i.e., using the first nine $W_N$ coefficients), which corresponded to an accuracy of $\sim 10^{-3}$. The interpolation based on expression (22) was unsatisfactory: the $\chi^2$ values obtained by treatment according to the power law (20) were abnormally large even for reasonable averaging intervals and gave no clear pattern with minima. The reason for this behavior is revealed by comparison of the $U_N$ coefficients (obtained by interpolation) to the exact values. As is seen from Fig. 10a, the difference is very large, making treatment by the power law practically impossible. Deviations increase by approximately the same law as those for the random errors, but the variation is rather smooth and is analogous for different $b_0$ values. It appears that these deviations can be compensated in a broad range of $b_0$ by optimization with respect to $\tilde{N}$.

This is really so and the region of optimum $\tilde{N}$ values can be determined without knowledge of the exact result. Figure 11 shows the behavior of $\chi^2$ in the interval of $20 \le N \le 40$ depending on $\tilde{N}$ for integer $b_0$ values. As is seen, small $\chi^2$ values are immediately obtained for $b_0 = 0, 1, 2, 3$ in the interval of $\tilde{N} = -(5.0\text{–}5.5)$. This is evidence that the error of $U_N$ can be compensated for all $b_0 \ge 0$, since greater $b_0$ correspond to still smaller errors (see Section 2.3). As is seen from Fig. 10b, deviations of the resulting $U_N$ for $\tilde{N} = -5.4$ from exact values for $b_0 \ge 0$ are in fact virtually indistinguishable.

The possibility of more refined optimization is based on the fact that the interpolation errors in formula (29) play the same role as do the high-order scaling corrections indicated by dots. As $\tilde{N}$ is changed, the interpolation errors smoothly vary and (for a certain $\tilde{N}$ value) become approximately compensated by the scaling corrections. This point can be detected by the maximum proximity of various estimates obtained for the $\alpha$ and $W_\infty$ values.

[8] Further increase in the number of optimization parameters seems to be inexpedient: this way may lead to absurd results. In particular, a large number of parameters allows imitation of a rapid convergence of the algorithm to an erroneous result. In the framework of the proposed approach, it is possible to ensure coincidence of four estimates of $\alpha$ and to obtain a zero value for estimated error.

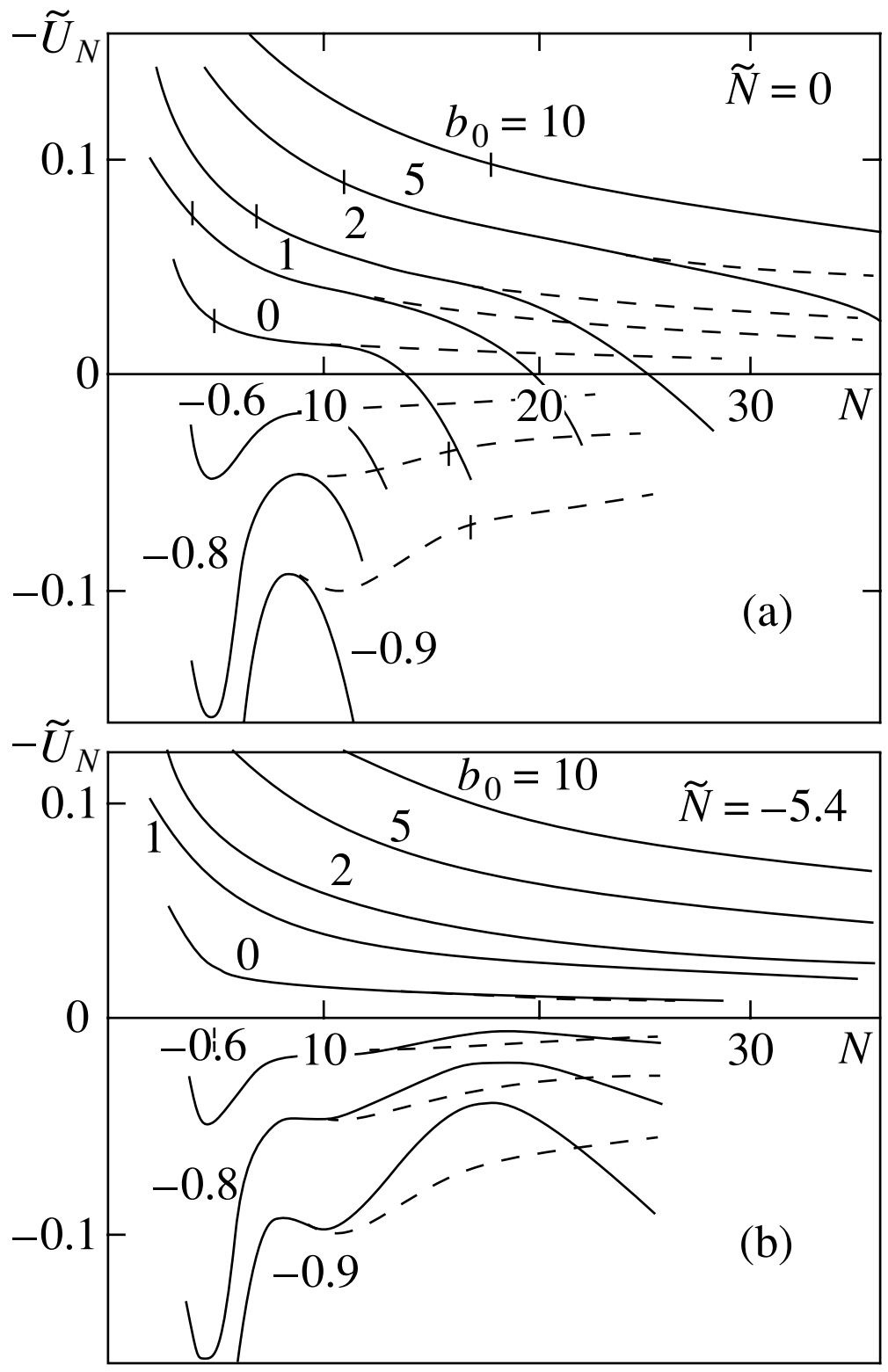

**Fig. 10.** Optimization of the interpolation procedure for an anharmonic oscillator: (a) a comparison of the $U_N$ values obtained by interpolation for $\tilde{N} = 0$ using the first nine $W_N$ coefficients (solid curves) to exact values (dashed curves); vertical bars indicate the $N$ values above which behavior of the exact $U_N$ values is visually indistinguishable from that according to the power law; (b) an analogous pattern after the optimization with respect to $\tilde{N}$ (for $\tilde{N} = -5.4$).

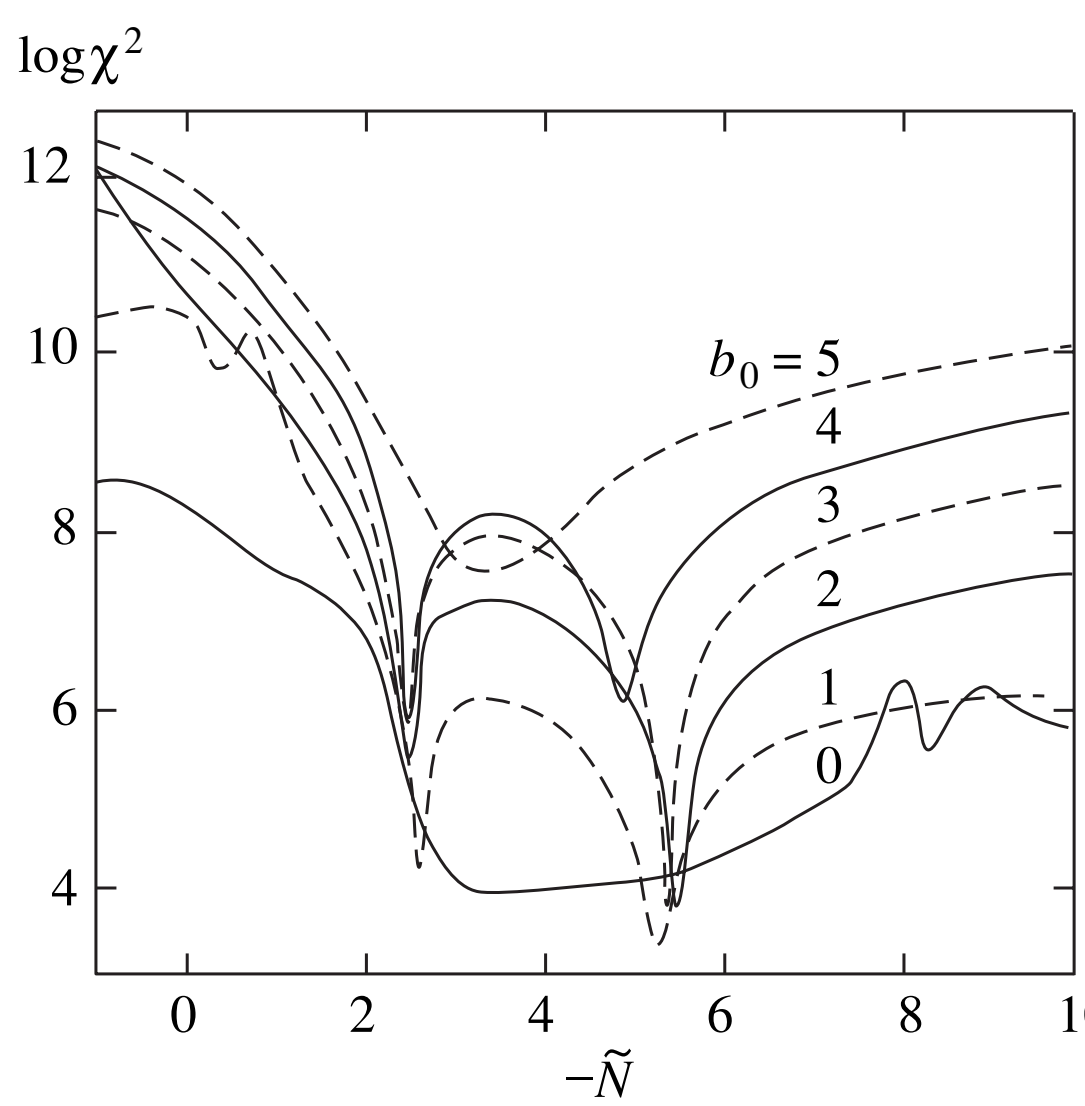

**Fig. 11.** The plots of $\chi^2$ versus $\tilde{N}$ for an anharmonic oscillator in the interval of $20 \le N \le 40$ at various fixed $b_0$ values.

A systematic treatment with determination of the $\alpha$ and $W_\infty$ values was carried out for $\tilde{N}$ in the interval from –5.0 to –5.6 at a step of 0.1. A "correct" pattern of $\chi^2$ minima was observed for $\tilde{N} = -5.5$, while for $\tilde{N} = -5.6$ the first minimum disappeared and for $\tilde{N} \ge -5.4$ it was split in two. The reason for this splitting is qualitatively evident: Figs. 10 and 11 show that, at a fixed $\tilde{N}$, there is a certain $b_0$ value for which the effect of the interpolation error upon $U_N$ is virtually compensated. This very $b_0$ corresponds to an "extra" minimum of $\chi^2$ in comparison with the pattern of Fig. 8. Since it is difficult to decide *a priori* which of the two minima is true, the estimates were obtained for both (and proved to be very close to each other).

The results of these numerical calculations are summarized in Table 3 and depicted in Fig. 12. The scatter of $\alpha$ and $W_\infty$ values allows the error to be evaluated by the order of magnitude. In order to obtain an "ideal" pattern according to Fig. 9b, the error interval should be expanded by a factor of 1.3 and 1.1 for $\alpha$ and $W_\infty$, respectively (dotted curves in Fig. 12). Then the values of $\alpha = 0.38$ and $W_\infty = 0.52$ (dashed curves in Fig. 12) are compatible with the results for all $\tilde{N}$. Selecting the $\tilde{N}$ values in each particular case so as to minimize the one-side error (as indicated by arrows in Fig. 12), we obtain the following estimates:

$$\alpha = 0.38 \pm 0.05, \quad W_\infty = 0.52 \pm 0.12. \tag{67}$$

A comparison to the set (57) shows that the error is estimated adequately, while the average values are somewhat displaced; the shift in $W_\infty$ is induced by the shift in $\alpha$.

## 7. SUMMING PERTURBATIVE SERIES FOR AN ARBITRARY $g$

When the amount of information concerning the $W_N$ coefficients suffices for restoring the $W(g)$ asymptotics as $g \longrightarrow \infty$, summing series (1) for an arbitrary $g$ encounters no problems: the coefficients $U_N$ for $N \lesssim 40$ are calculated by formula (19) and the subsequent terms can be obtained according to the $U_\infty N^{\alpha-1}$ asymptotics, so that all coefficients of the converging series (9) are known. The summation error is determined by the accuracy of restoring the asymptotics,

$$\Delta_{as} = \left.\frac{\delta U_N}{U_N}\right|_{N \gg 1} = \frac{\delta U_\infty}{U_\infty} + \delta\alpha \ln N, \tag{68}$$

which varies logarithmically with $N$ and can be considered as constant with a restricted interval. Introducing a characteristic scale $N_c$ on which the relative error is comparable with $\Delta_{as}$ and using the approximation

$$\frac{\delta U_N}{U_N} = \begin{cases} 0, & N < N_c \\ \Delta_{as}, & N \ge N_c, \end{cases} \tag{69}$$

we obtain for $ag \gg 1$

$$\delta B(g) = \sum_{N=N_c}^{\infty} \Delta_{as} U_N \exp\left(-\frac{N_c}{ag}\right) = \begin{cases} \Delta_{as} B(g), & ag \gg N_c \\ \Delta_{as} U_{N_c} ag \exp(-N_c/ag), & ag \ll N_c. \end{cases} \tag{70}$$

**Table 3.** Asymptotic parameters for an anharmonic oscillator obtained by the interpolation with $L_0 = 1$, $L = 9$ (the values in parentheses for $\tilde{N} = -5.6$ were estimated at the point $b_0 = 2.20$ where the first $\chi^2$ minimum disappears)

| Estimates based on | α for $\tilde{N}$ | | | | | | |
|---|---|---|---|---|---|---|---|
| | –5.0 | –5.1 | –5.2 | –5.3 | –5.4 | –5.5 | –5.6 |
| First $\chi^2$ minimum | 0.398<br>0.476 | 0.396<br>0.452 | 0.393<br>0.422 | 0.390<br>0.399 | 0.385<br>0.384 | 0.378 | (0.373) |
| Second $\chi^2$ minimum | 0.50 | 0.47 | 0.42 | 0.37 | 0.33 | 0.29 | 0.34 |
| $U_\infty$ changing sign | 0.585 | 0.535 | 0.485 | 0.445 | 0.405 | 0.365 | 0.335 |
| $U_\infty(b_0)$ linearization | 0.495 | 0.445 | 0.40 | 0.36 | 0.32 | 0.29 | 0.26 |
| | $W_\infty$ for $\tilde{N}$ | | | | | | |
| | –5.0 | –5.1 | –5.2 | –5.3 | –5.4 | –5.5 | –5.6 |
| First $\chi^2$ minimum | 0.490<br>0.356 | 0.495<br>0.390 | 0.500<br>0.440 | 0.505<br>0.487 | 0.513<br>0.517 | 0.529 | (0.540) |
| $U_\infty(b_0)$ slope | 0.226<br>0.502 | 0.290<br>0.538 | 0.373<br>0.568 | 0.463<br>0.698 | 0.572<br>0.885 | 0.675<br>1.09 | 0.712<br>0.953 |

Substituting these expressions into (8) and using the steepest descent method for $ag \ll N_c$, we obtain

$$\frac{\delta W(g)}{W(g)} \sim \begin{cases} \Delta_{as}, & ag \gtrsim N_c \\ \Delta_{as} \exp\{-2(N_c/ag)^{1/2}\}, & ag \lesssim N_c \end{cases} \tag{71}$$

(where some preexponential factors are omitted for clarity). For negative α, the results for $ag \gg N_c$ are somewhat different. In particular, for $-1 < \alpha < 0$ we obtain $\delta W(g) = \Delta_{as}(W(g) - W(g_c))$, where $ag_c \sim N_c$. A natural scale for $N_c$ is provided by the middle of the working interval $(N_{min}, N_{max})$, that is, $N_c \approx 30$; however, deviations from this value may be quite large because the corresponding equality holds in fact on the logarithmic scale ($\ln N_c \approx \ln 30$). In practice, approximation (69) with a constant $N_c$ is expedient only for large $g$. In the general case, estimate (71) is valid with an effective $N_c$ value, which is determined by the number $N$ of the maximum term $\delta U_N u^N$ in the series for $\delta B(u)$ (for small $g$, this value is close to $L + 1$, e.g., to the number of the first unknown coefficient $W_N$).

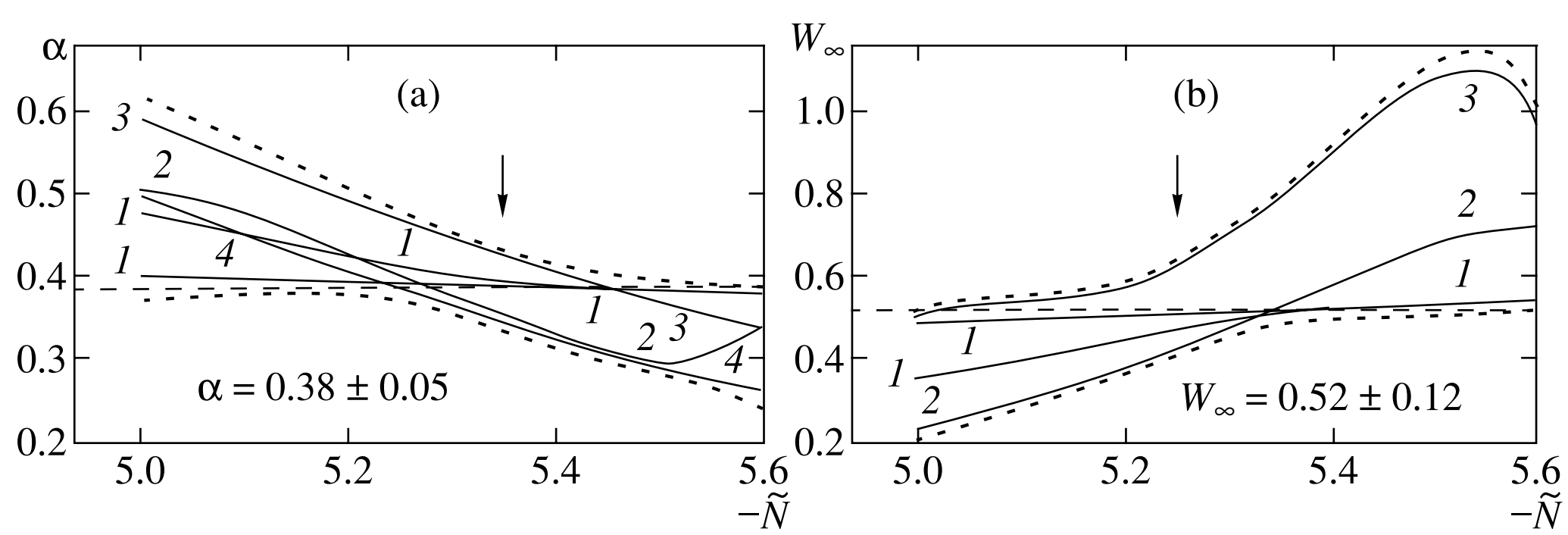

**Fig. 12.** The plots of α and $W_\infty$ values estimated for an anharmonic oscillator by various methods (see Section 3): (a) α estimates based on the (*1*) first $\chi^2$ minimum, (*2*) second $\chi^2$ minimum, (*3*) $U_\infty$ changing sign, and (*4*) $U_\infty(b_0)$ linearization; (b) $W_\infty$ estimates based on the (*1*) first $\chi^2$ minimum and (*2*, *3*) $U_\infty(b_0)$ slope (upper and lower bounds, respectively). Small-dash lines indicate the error interval expanded by a factor of 1.3 and 1.1 for α and $W_\infty$ values, respectively.

**Table 4.** Comparative data for the exact integral (32) with $n = 1$ and the results obtained by summing the perturbative series

| $g$ | $W(g) \times 10$ | | | |
|---|---|---|---|---|
| | Exact value | Summing with exact $W_N$ | Summing upon interpolation with $L_0 = 1$, $L = 5$ | Summing upon interpolation with $L_0 = 1$, $L = 1$ |
| 1 | 6.842134 | 6.842135 | 6.842134 | 6.8436 |
| 2 | 6.183453 | 6.183454 | 6.183452 | 6.1867 |
| 4 | 5.497111 | 5.497110 | 5.497105 | 5.5034 |
| 8 | 4.820615 | 4.820608 | 4.820594 | 4.832 |
| 16 | 4.181699 | 4.181669 | 4.181637 | 4.200 |
| 32 | 3.597297 | 3.59720 | 3.59714 | 3.624 |
| 64 | 3.075230 | 3.07500 | 3.07490 | 3.113 |
| 128 | 2.616802 | 2.61633 | 2.61617 | 2.668 |
| 256 | 2.219222 | 2.2184 | 2.2182 | 2.285 |
| 512 | 1.877472 | 1.8761 | 1.8758 | 1.959 |
| 1024 | 1.585578 | 1.5835 | 1.5831 | 1.68 |
| $g \longrightarrow \infty$ | $9.064g^{-0.25}$ | $8.95g^{-0.247}$ | $8.95g^{-0.247}$ | $9.12g^{-0.269}$ |

**Table 5.** Comparative data for the exact ground state energy $E_0(g)$ of an anharmonic oscillator and the results obtained by summing the perturbative series (the $2E_0(g)$ and $2g$ values are given in order to provide for the correspondence with the data reported in most other papers using a different normalization)

| $2g$ | $2E_0(g)$ | | |
|---|---|---|---|
| | Exact value | Summing with exact $W_N$ ($b_0 = 1.30$) | Summing upon interpolation with $L_0 = 1$, $L = 9$ ($\tilde{N} = -5.3$, $b_0 = 3.55$) |
| 0.5 | 1.241854 | 1.241854 | 1.241857 |
| 1 | 1.392352 | 1.392352 | 1.392396 |
| 2 | 1.607541 | 1.607545 | 1.60790 |
| 3 | 1.769589 | 1.769605 | 1.7706 |
| 4 | 1.903137 | 1.903178 | 1.9051 |
| 5 | 2.018341 | 2.018418 | 2.0214 |
| 10 | 2.449174 | 2.44961 | 2.4599 |
| 20 | 3.009945 | 3.0117 | 3.040 |
| 50 | 4.003993 | 4.0115 | 4.096 |
| 100 | 4.999418 | 5.018 | 5.19 |
| $g \longrightarrow \infty$ | $2 \times 0.668g^{1/3}$ | $2 \times 0.602g^{0.349}$ | $2 \times 0.511g^{0.387}$ |

Table 4 presents the results of calculations for the zero-dimensional case. Here, the first column gives the exact values of integral (32) with $n = 1$, while the columns from second to fourth present the results of summation obtained using exact $W_N$ coefficients and interpolated values (with $L_0 = 1$, $L = 5$ or $L_0 = 1$, $L = 1$), respectively. In each case, the calculations were performed for $b_0$ corresponding to the first $\chi^2$ minimum. A comparison to (71) indicates that $N_c \sim 200$ for the second and third columns and $N_c \sim 10$ for the fourth column.

Table 5 presents the analogous data for an anharmonic oscillator. Here, the first column gives the exact $E_0(g)$ values taken from [24], while the second and third columns present the results of summation obtained using exact $W_N$ coefficients and interpolated values (with $L_0 = 1$, $L = 9$), respectively. In this case, the estimates give $N_c \sim 200$ for the second column and about 50 for the third column.

Information concerning the $W(g)$ asymptotics can also be taken into account within the framework of the standard conform-Borel procedure (Section 2.1) by interpolating the $U_N$ coefficients (with the known asymptotics (17)) calculated using formula (11). For approximation (69), we obtain by analogy with (71)

$$\frac{\delta W(g)}{W(g)} \sim \begin{cases} \Delta_{as}, & ag \gtrsim N_c^2 \\ \Delta_{as} \exp\{-3(N_c^2/ag)^{1/3}\}, & ag \lesssim N_c^2. \end{cases} \quad (72)$$

This procedure is preferred in the case of sufficiently small $g$ values (when $N_c$ is close to $L + 1$), leading to smaller errors as compared to those obtained for (71). For greater $g$, the attaining of $N_c$ values indicated above seems to be impossible.

According to the standard procedure of calculating the critical indices [6], the second series (9) is truncated on the $L$th term that corresponds to the error given by (72) with $N_c = L + 1$ and $\Delta_{as} \sim 1$. In the three-dimensional case, a large number of expansion coefficients are known (for $L = 6$). These values are well matched with (2), which gives hope for restoring the asymptotics of scaling functions with an accuracy of $\Delta_{as} \sim 10^{-2}$ and for increasing $N_c$ at the expense of interpolation. Thus, it is apparently possible to increase the accuracy of calculation of the critical indices by two–three orders of magnitude even for the currently available information. Using the modified conformal mapping may lead to a further increase in the accuracy, provided that the scale of $N_c \gtrsim 20$ would be accessible in the corresponding region of $ag \sim 0.2$.

## 8. THE $\varphi^4$ THEORY

### *8.1. Restoration of the Gell-Mann–Low Function*

Now let us turn to a real physical problem of restoring the Gell-Mann–Low function in the $\varphi^4$ theory, considering $\beta(g)$ as $W(g)$ and proceeding from the information contained in relationships (5) and (6).

The interpolation was based on formula (63) with an optimum value of $\tilde{b} = 4$. Figure 13 presents the plots of $\chi^2(\tilde{N})$ versus $\tilde{N}$ calculated in the interval $20 \le N \le 40$ for several fixed $b_0$ values. As is seen, promising results can be expected for $\tilde{N}$ values close to zero, where the curves obtained at $b_0 = -1, 0, -1$ and 2 exhibit sharp minima. The interval $-0.5 \le \tilde{N} \le 0.5$ was studied in more detail.

Figure 14 shows the behavior of the coefficients $\tilde{U}_N = U_N\Gamma(b_0 + 2)$ in the case of a nearly optimum interpolation with $\tilde{N} = 0$. If the curves for $b_0 \gg 1$ and $b_0 \approx -2$ (attaining the asymptote with delay) are rejected, the data for large $N$ asymptotically tend to a constant level, which correspond to a critical index $\alpha$ close to unity. This conclusion is consistent with the position of the second $\chi^2$ minimum and with the change of sign in $U_\infty$ (Fig. 15). A clear pattern with $\chi^2$minima was observed for $\tilde{N} \le 0.2$; when the $\tilde{N}$ value increased, the first $\chi^2$ minimum approached to and eventually merged with the second minimum. For this reason, no estimates using the first minimum could be obtained for $\tilde{N} \ge 0.3$.

The results of determining the $\alpha$ and $W_\infty$ values are presented in Table 6 and Fig 16. The ideal pattern for $\alpha$, corresponding to Fig. 9b, is obtained upon expanding the error interval by a factor of two (dashed lines in Fig. 16a), after which the value of $\alpha = 0.96$ is compatible with the results for all $\tilde{N}$. In the fixed interval of $20 \le N < 40$, all four estimates of $\alpha$ coincide for $\tilde{N} = -0.12$ on an accuracy level of $10^{-3}$; the main uncertainty is related to a weak dependence on the averaging interval. With an allowance for the double error, we finally obtain

$$\alpha = 0.96 \pm 0.01. \tag{73}$$

For $W_\infty$ (Fig. 16b), the ideal pattern is obtained immediately and the corresponding value of $W_\infty = 7.4$ is com-

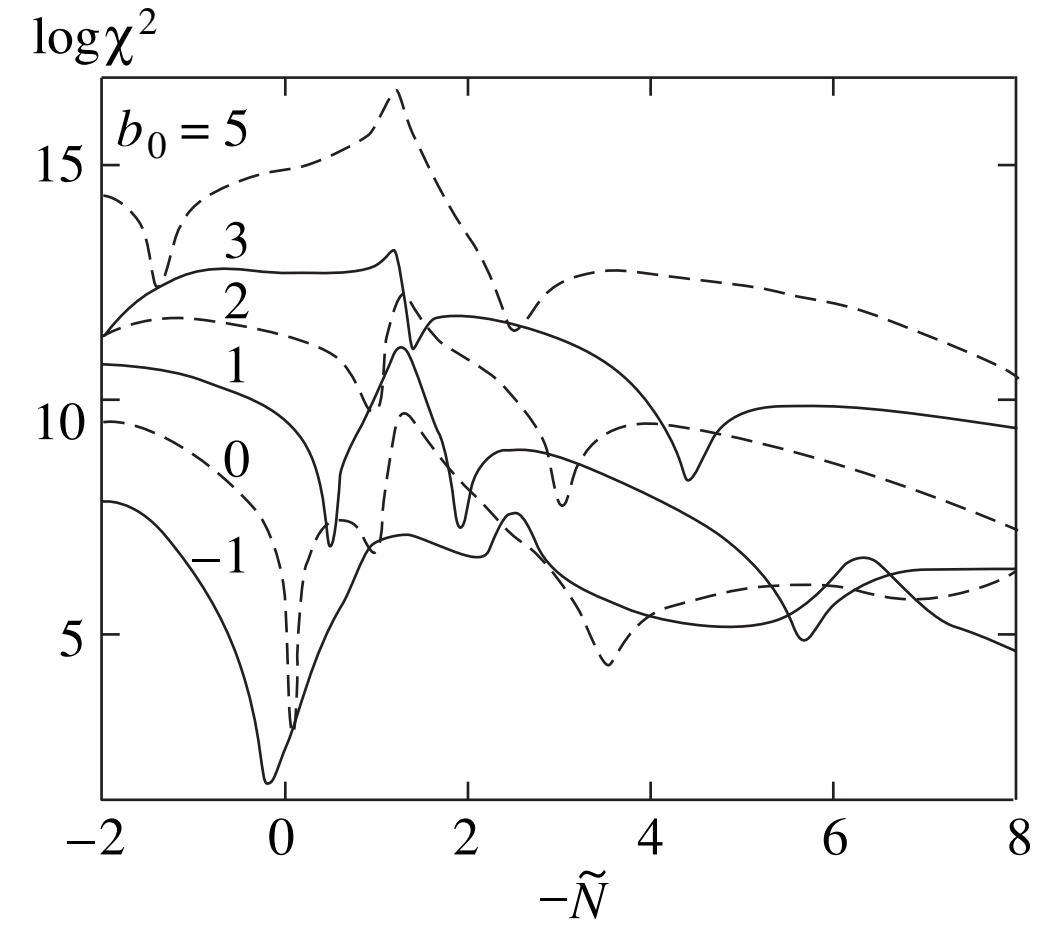

**Fig. 13.** The plots of $\chi^2$ versus $\tilde{N}$ for the $\varphi^4$ theory in the interval of $20 \le N \le 40$ at various fixed $b_0$ values.

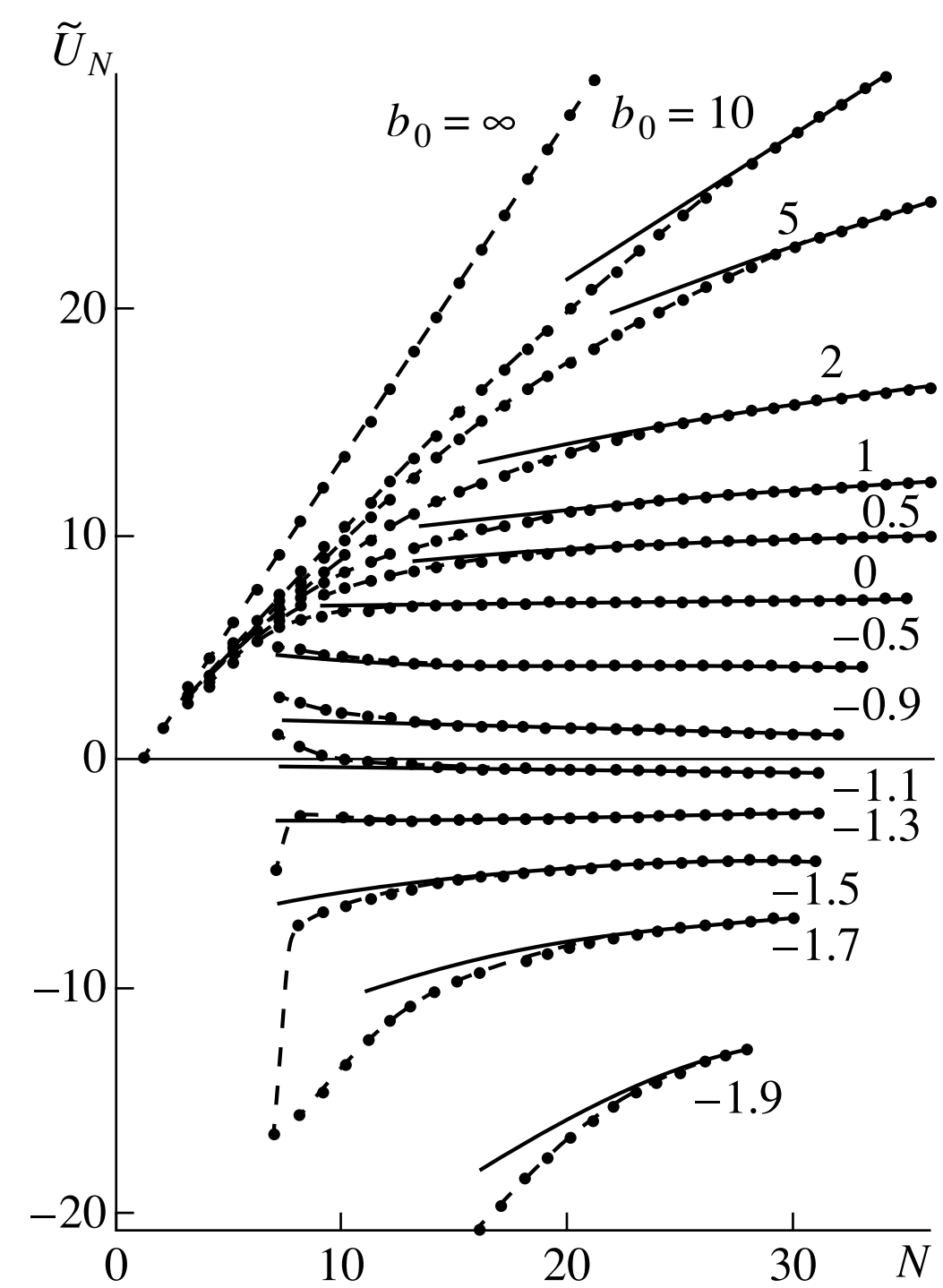

**Fig. 14.** The plots of $\tilde{U}_N = U_N\Gamma(b_0 + 2)$ versus $N$ for various $b_0$ (points and dashed curves) and the results of treatment according to the power law (solid curves) for the $\varphi^4$ theory. The calculations were performed using a nearly optimum interpolation with $\tilde{b} = 4$, $\tilde{N} = 0$.

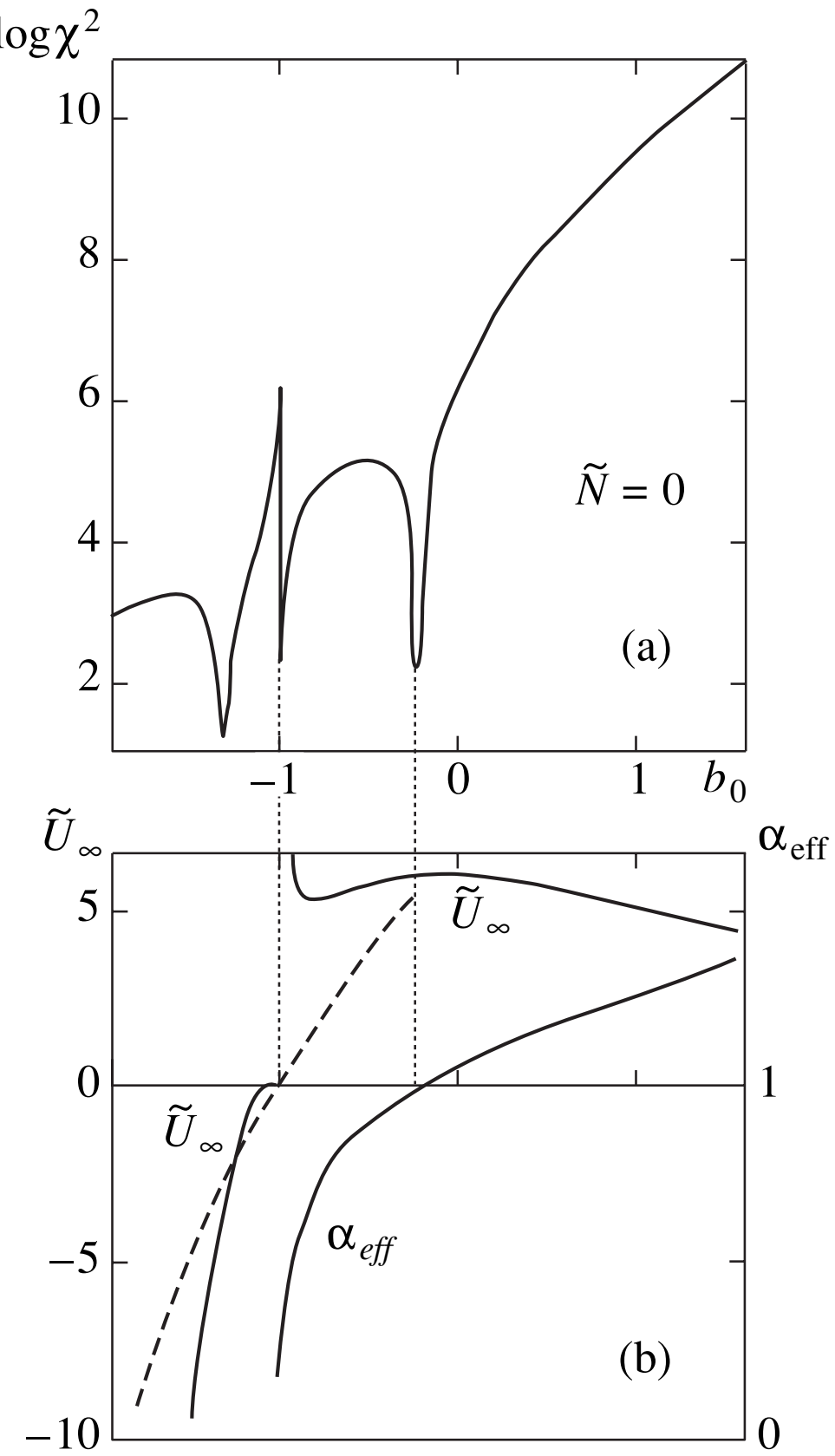

**Fig. 15.** (a) The pattern of minima in $\chi^2$ for the $\varphi^4$ theory in the averaging interval of $20 \le N \le 40$. (b) The plots $\alpha_{\rm eff}$ and $\tilde{U}_\infty$ versus $b_0$ for $\tilde{N} = 0$. The dashed curve shows the $U_\infty(b_0)$ curve for fixed $\alpha = 1$.

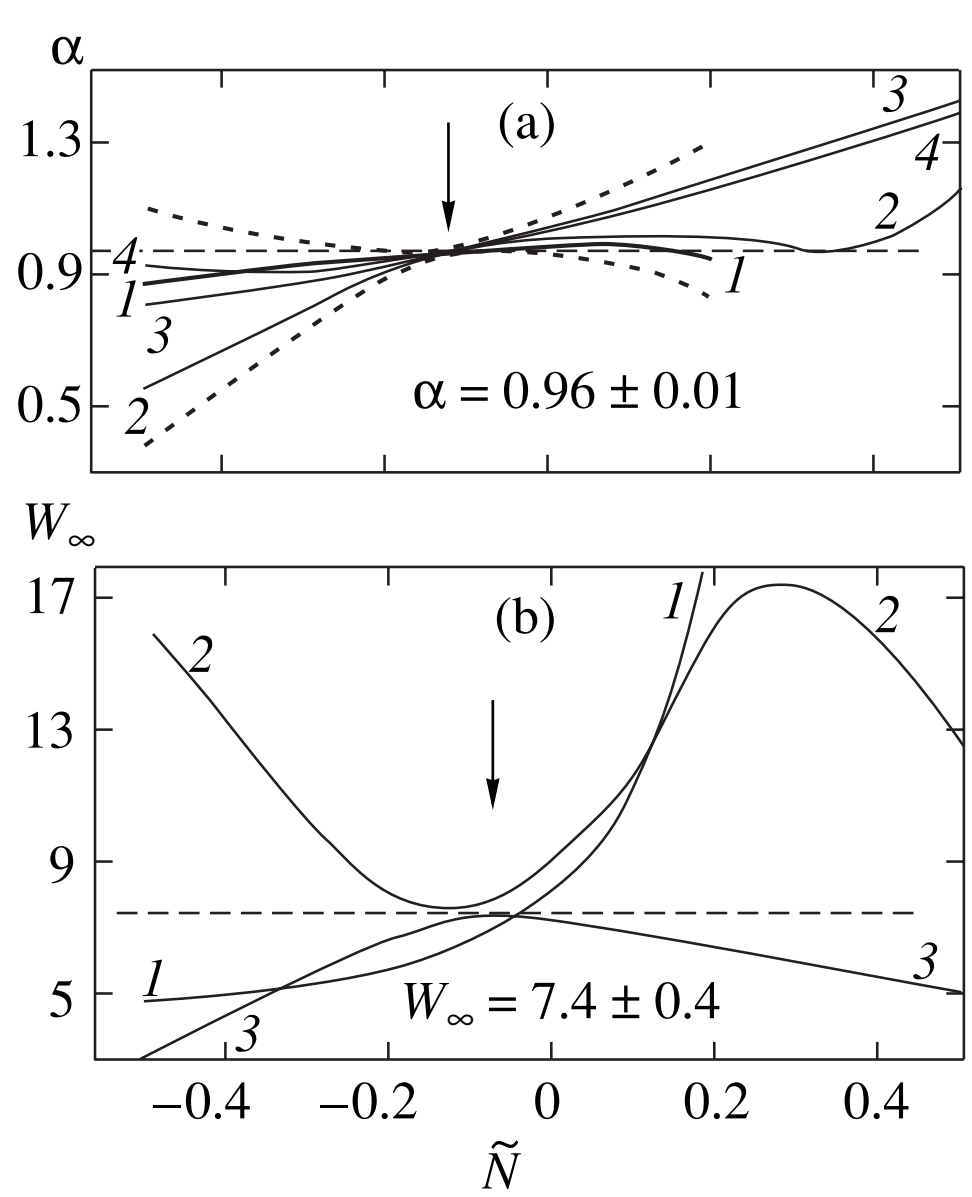

**Fig. 16.** The plots of various (a) $\alpha$ and (b) $W_\infty$ estimates versus $b_0$ for the $\varphi^4$ theory. The notations are the same as in Fig. 12. Small-dash lines indicate the error interval for $\alpha$ expanded by a factor of two.

patible with all data. Here, the one-side error is minimum at $\tilde{N} = -0.08$, which yields

$$W_\infty = 7.4 \pm 0.4. \quad (74)$$

Correctness of the optimization with respect to $\tilde{b}$ (which was carried out in Section 6.2 in somewhat heuristic manner) can be demonstrated now. For an optimum value of $\tilde{N} = -0.12$ and the $\tilde{b}$ value varied in an interval from 0 to 6, a clear pattern of $\chi^2$ minima was obtained in the middle of the interval. On approaching the boundaries, the first $\chi^2$ minimum approached to and merged with the second minimum exactly as it was observed on increasing $\tilde{N}$. These corresponding results for $\alpha$ and $W_\infty$ are presented in Fig. 17; expanding the error interval by a factor of 2 and 1.1 for $\alpha$ and $W_\infty$, respectively, makes the values (73) and (74) compatible with almost all data (except for a narrow interval at $\tilde{b} = 5.5$, where the proximity of all estimates is obviously accidental. As is seen, the minimum errors also agree with (73) and (74).

Summation of the perturbative series for the Gell-Mann–Low function at finite $g$ values was performed using a procedure analogous to that described in Section 7. The accuracy was evaluated by variation with respect to $b_0$ and $\tilde{N}$. The variation with respect to $b_0$ gave a markedly greater $N_c$ values and allowed the $W(g)$ asymptotics to be modified without significantly affecting the results for $g \sim 1$. On varying the $\tilde{N}$ value, with $b_0$ adjusted so as to maintain a constant value of $\alpha = 0.96$, the most probable value of $W_\infty = 7.4$ is obtained for $\tilde{N} = -0.067$; the uncertainty range indicated in (74) corresponds to the interval $-0.09 \le \tilde{N} \le -0.05$. Table 7 lists the data for $\tilde{N} = -0.067$, with the error estimated by comparison to the results for $\tilde{N} = -0.05$ and $-0.09$. Note that asymptote (7) is attained rather slowly, the deviation amounting to about 15% even for $g = 100$.

Figure 18 presents a comparison of the results obtained for $g \le 20$ to the data reported by other researchers.

### *8.2. The Possibility of Logarithmic Branching*

Since the value of $\alpha$ differs only slightly from unity, a question arises as to whether the accuracy is sufficient to consider this deviation significant. Formally speaking, this is really so because the error was estimated objectively and there is no ground to expect it to be significantly understated. Nevertheless, the possibility that the equality $\alpha = 1$ is strict is not excluded, since asymptotics (7) may contain logarithmic corrections of the type

$$W(g) = W_\infty g^\alpha (\ln g)^{-\gamma}, \quad g \longrightarrow \infty. \quad (75)$$

For $\gamma > 0$, these corrections may inspire a small decrease in $\alpha$. In this case, formula (20) contains an additional factor $(\ln N)^{-\gamma}$ with unchanged $W_\infty$, so that the results for $U_N$ can be treated according to Eq. (75) with the parameters

$$\alpha = 1, \quad \gamma \approx 0.14, \quad W_\infty \approx 7.7 \qquad (76)$$

without any increase in $\chi^2$. Actually, the possibility of such a logarithmic branching seems to be quite probable for the following reasons.

**Table 6.** Asymptotic parameters for the $\varphi^4$ theory obtained for $\tilde{b}_0 = 4$ and various $\tilde{N}$ values by the interpolation with $L_0 = 2$, $L = 5$

| Estimates based on | $\alpha$ for $\tilde{N}$ | | | | | | | | | |
|---|---|---|---|---|---|---|---|---|---|---|
| | –0.5 | –0.3 | –0.2 | –0.12 | –0.1 | 0.0 | 0.1 | 0.2 | 0.3 | 0.5 |
| First $\chi^2$ minimum | 0.863 | 0.920 | 0.945 | 0.962 ±0.005 | 0.964 | 0.975 | 0.974 | 0.931 | – | – |
| Second $\chi^2$ minimum | 0.54 | 0.78 | 0.90 | 0.960 | 0.970 | 1.00 | 1.01 | 1.01 | 0.97 | 1.16 |
| $U_\infty$ changing sign | 0.795 | 0.865 | 0.915 | 0.960 –0.961 | 0.973 | 1.035 | 1.105 | 1.175 | 1.255 | 1.415 |
| $U_\infty(b_0)$ linearization | 0.907 | 0.90 | 0.929 | 0.961 ±0.001 | 0.971 | 1.022 | 1.082 | 1.147 | 1.218 | 1.371 |
| | $W_\infty$ for $\tilde{N}$ | | | | | | | | | |
| | –0.5 | –0.3 | –0.2 | –0.12 | –0.1 | 0.0 | 0.1 | 0.2 | 0.3 | 0.5 |
| First $\chi^2$ minimum | 4.67 | 5.22 | 5.75 | 6.36 ±0.16 | 6.63 | 8.26 | 11.82 | 30.9 | – | – |
| $U_\infty(b_0)$ | 3.02 | 5.58 | 6.55 | 7.35 | 7.34 | 7.18 | 6.78 | 6.45 | 5.91 | 5.05 |
| slope | 15.9 | 10.0 | 7.85 | 7.55 | 7.61 | 9.07 | 11.3 | 16.5 | 17.3 | 12.3 |

**1.** It can be ascertained that the logarithmic branching in the case of strict equality $\alpha = 1$ is unavoidable. Indeed, let us write series (1) in the form of the Sommerfeld–Watson integral [2, 13]:

$$W(g) = \sum_{N=N_0}^{\infty} W_N(-g)^N = -\frac{1}{2i}\oint_C dz \frac{\mathcal{W}(z)}{\sin \pi z} g^z, \qquad (77)$$

where $\mathcal{W}(z)$ is the analytical continuation of $W_N$ onto the complex plane ($\mathcal{W}(N) = W_N$) and $C$ is the contour containing the points $N_0$, $N_0 + 1$, $N_0 + 2$, … (Fig. 19). If $z = \alpha$ is the extreme right-hand singularity of $\mathcal{W}(z)/\sin \pi z$, we can modify the contour into the position $C'$ and show that this singularity determines the behavior of $W(g)$ as $g \longrightarrow \infty$. The purely power law (7) corresponds to the presence of a simple pole at $z = \alpha$, while the law described by Eq. (75) corresponds to a singularity of the $(z - \alpha)^{\gamma - 1}$ type.[9]

[9] It is clear from the above considerations that the assumption of analyticity of the coefficient function on the real axis for $N \geq N_0$, which is necessary for interpolation, is confirmed in all cases by the results obtained.

Note that the first term $\beta_0$ is absent in the expansion of the $\beta$ function (5) simply by its definition, while vanishing of the next coefficient $\beta_1$ is accidental. Indeed, in the $(4 - \epsilon)$-dimensional $\varphi^4$ theory, the latter term is nonzero and has a magnitude on the order of $\epsilon$; accordingly, $\mathcal{W}(1) \sim \epsilon$. The limiting transition $\epsilon \longrightarrow 0$ shows that, in the four-dimensional case, $\mathcal{W}(1) = 0$ and a simple pole cannot take place at $\alpha = 1$. If the function $\mathcal{W}$ tends to zero as $z \longrightarrow 1$ by the law $\mathcal{W}(z) = \omega_0(z-1)^\gamma$, then

$$\beta(g) = \frac{\omega_0}{\Gamma(1-\gamma)} g(\ln g)^{-\gamma}, \quad g \longrightarrow \infty \qquad (78)$$

and the positive definiteness of $\gamma$ has a quite clear origin.

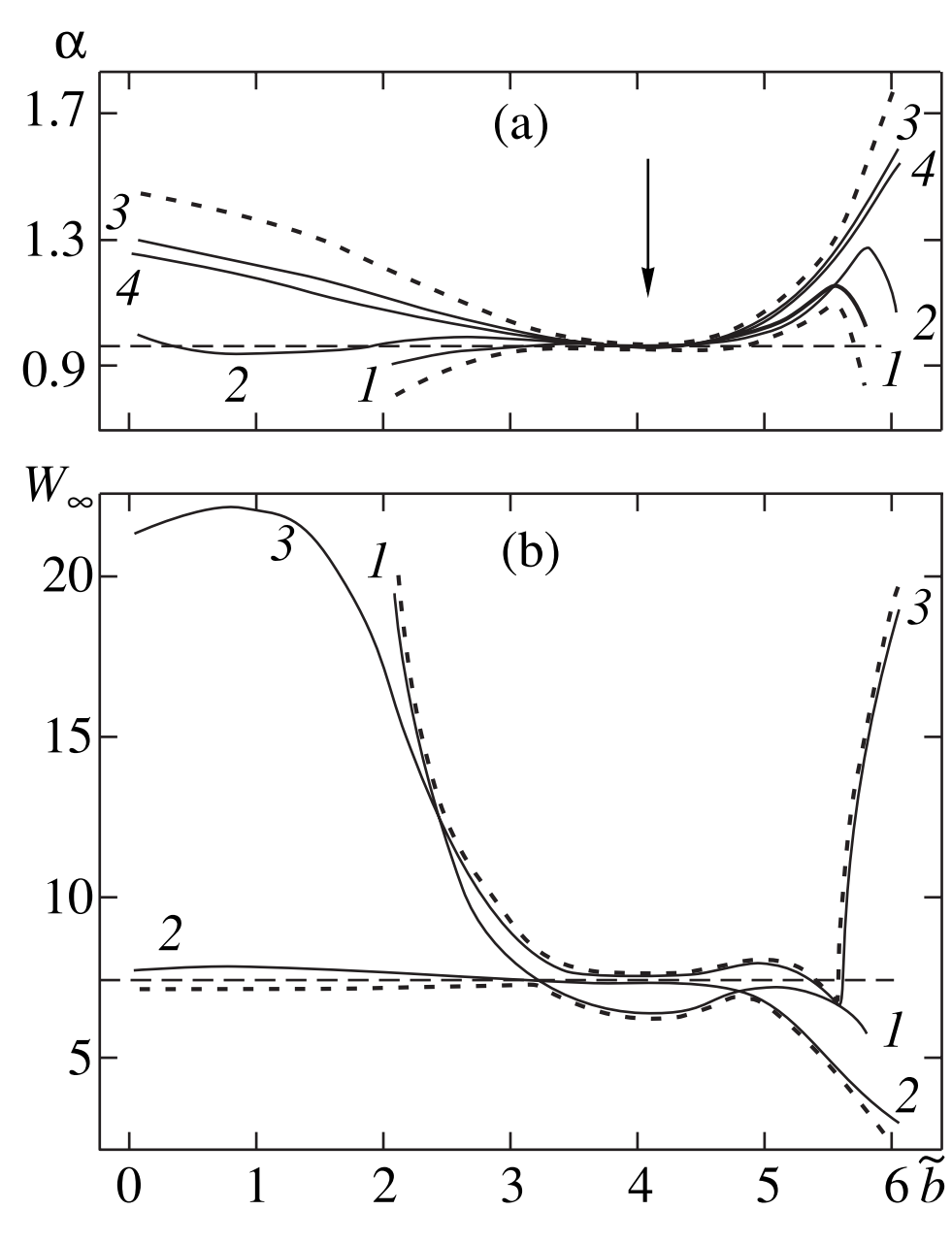

**Fig. 17.** The plots of various (a) $\alpha$ and (b) $W_\infty$ estimates versus $\tilde{b}$ for the $\varphi^4$ theory. The notations are the same as in Fig. 12. Small-dash lines indicate the error interval expanded by a factor of 2 and 1.1 for $\alpha$ and $W_\infty$ values, respectively.

**Table 7.** The Gell-Mann–Low function for the $\varphi^4$ theory (values in parentheses indicate the error estimated in units of the last decimal digit)

| $g$ | $\beta(g)$ | $g$ | $\beta(g)$ |
|---|---|---|---|
| 0.2 | 0.04993(2) | 30 | 138.7(50) |
| 0.4 | 0.18518(26) | 40 | 193.2(75) |
| 0.6 | 0.3939(10) | 50 | 248.3(100) |
| 0.8 | 0.6667(27) | 60 | 303.9(127) |
| 1 | 0.9952(51) | 70 | 359.7(155) |
| 2 | 3.272(33) | 80 | 415.6(182) |
| 3 | 6.278(85) | 90 | 471.7(212) |
| 4 | 9.758(157) | 100 | 527.7(240) |
| 5 | 13.57(25) | 150 | 808.1(389) |
| 6 | 17.64(36) | 200 | 1087(54) |
| 7 | 21.90(47) | 250 | 1366(70) |
| 8 | 26.32(60) | 300 | 1644(86) |
| 9 | 30.87(75) | 350 | 1920(101) |
| 10 | 35.53(90) | 400 | 2196(127) |
| 15 | 59.95(175) | 450 | 2471(133) |
| 20 | 85.59(275) | 500 | 2745(149) |
| 25 | 111.9(38) | $g \longrightarrow \infty$ | $7.41g^{0.96}$ |

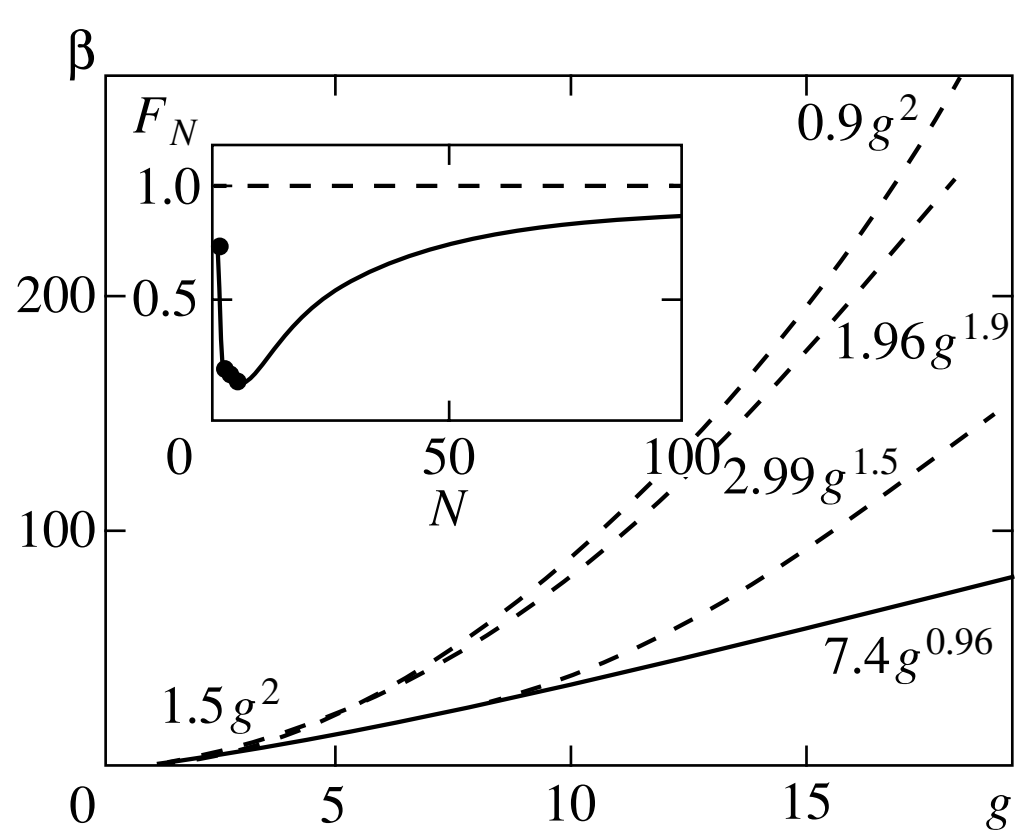

**Fig. 18.** A comparison of the Gell-Mann–Low function for the $\varphi^4$ theory calculated in this work (solid curve) to the results reported by other researchers (dashed curves top to bottom corresponding to [12, 13, 14], respectively). The inset shows a reduced coefficient function (in this scale, differences between the data obtained using various interpolation methods are insignificant).

**2.** Lipatov [29] considered the class of field theories (generalizing the four-dimensional $\varphi^4$ theory) with a nonlinearity of the $\varphi^n$ type and a space dimension of $d = 2n/(n-2)$, for which a logarithmic situation takes place. For all such theories, $\beta_1 = 0$; however, this coefficient differs from zero when $d$ decreases. Therefore, $\mathcal{W}(1) = 0$ by analogy with the cases considered above. In the limit $n \longrightarrow \infty$, the Gell-Mann–Low function is exactly calculated [29] and the extreme right-hand singularity of $\mathcal{W}(z)$ has the form of $(z-1)^{3/2}$, which leads to asymptotics of the type $\beta(g) \propto g(\ln g)^{-3/2}$. From the continuity considerations, we may expect for large but finite $n$ values that a nonanalytical zero of the type $(z-1)^\gamma$ is retained and the singularity at $z = 1$ is still the extreme right-hand one. Therefore, asymptotics (78) is natural for such field theories and it is not surprising that it may be retained up to $n = 4$. Note that $W_\infty$ is negative when $n \longrightarrow \infty$, so that the Gell-Mann–Low function possesses a zero; a direct extrapolation of the results to $n = 4$ leads to an analogous conclusion for the $\varphi^4$ theory [29]. In fact, with this extrapolation we must take into account that the index $\gamma$ changes from 3/2 to small values such as in (76); then the change in sign of the asymptotics naturally takes place according to (78) at $\gamma = 1$. The positiveness of $\omega_0$ follows from the matching of $\mathcal{W}(2) \sim \omega_0$ and the positiveness of $\beta_2$ [29].

Anyhow, we have to select between two possibilities: (i) a purely power law (7) with a critical index $\alpha$ slightly below unity and (ii) an asymptotics of the type (78) with $\gamma > 0$. In both cases, the $\varphi^4$ theory turns out to be self-consistent.

### *8.3. On the Results Obtained in [12, 13]*

The curves in Fig. 14 display for $N < 10$ a linear portion where $\tilde{U}_N \approx 1.1(N-1)$, which is stable with respect to changes both in $b_0$ and in the extrapolation procedure. This region might be considered as a true asymptotics for $\tilde{U}_N$ (assuming the results for $N > 10$ to be the interpolation artifacts), corresponding to the dependence $\beta(g) \approx 1.1g^2$, which is close to the result obtained in [12, 13].

In fact, stability of the above region has a different origin. This behavior is related to a characteristic "trough" in the reduced coefficient function $F_N$ at $N \lesssim 10$ (see the inset in Fig. 18). Modeling this trough by assuming $F_3 = F_4 = \ldots = F_{10} = 0$ and taking into account Eqs. (19) and (22), we obtain

$$\tilde{U}_N = c\Gamma(b_0+2) \times \sum_{K=1}^{N} F_K(-1)^K \frac{\Gamma(K+b)}{\Gamma(K+b_0)} C_{N-1}^{K-1}. \quad (79)$$

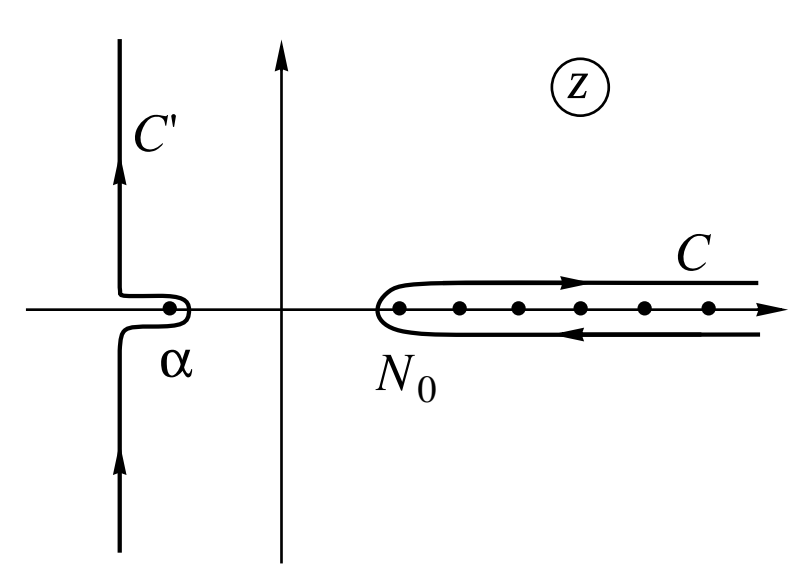

**Fig. 19.** Integration contour for Eq. (77).

From this we obtain for $N \le 10$ and all $b_0$ the result $\tilde{U}_N = 1.5(N-1)$, which is determined by the first nonvanishing coefficient $F_2$ (see the curve for $b_0 = \infty$ in Fig. 14) and is close to the real situation. For the β function, this result implies that a single-loop law $1.5g^2$ is valid up to $g \sim 10$.

Upon modeling the trough in $F_N$ more precisely by assuming $F_3 = F_4 = \ldots = F_{10} = \epsilon$ and using (26) we obtain, in the case when the ratio of gamma functions in (79) reduces to a polynomial for $b_0 = b - p$ with an integer $p$ and $N$ in the interval $p + 2 \le N \le 10$,

$$\tilde{U}_N = W_2\left\{\left(1-\frac{\epsilon}{F_2}\right)(N-1) + \frac{\epsilon}{F_2}\frac{1+b_0}{1+b}\right\}. \quad (80)$$

This result indicates that the linear slope varies but remains independent of $b_0$. More complicated calculations show that Eq. (80) is valid for arbitrary $b_0$ to within corrections on the order of $\epsilon/(N+b_0)^{b+1}$; for $\epsilon = 0.2$ (see Fig. 18), we obtain $\tilde{U}_N = 1.1(N-1) + \text{const}$, where the last constant depends on $b_0$ but does not exceed a few tenths in the interval $0 < b_0 < 10$. Thus, a notion of the quadratic law with modified coefficient $\beta(g) = 1.5(1-\epsilon/F_2)g^2$ is really meaningful in the interval $1 \lesssim g \lesssim 10$ but is a consequence of the trough in $F_N$.[10] The limited width of the trough indicates that this law is not related to a real asymptotics (whatever it is).

The above considerations clearly indicate that the result obtained in [12, 13] is by no means a computational error and objectively reflects the behavior of the β function for $g \lesssim 10$. This result is unavoidably obtained upon summing a series with a small number of expansion coefficients, since no other portion obeying the power law can be found in Fig. 14 for $N < 7$ (the points on the curves for $b_0 < 0$ are omitted for clarity, because their sharp oscillations would overload the pattern).

### *8.4. The Question of "Triviality" of the $\varphi^4$ Theory*

The situation when the β function possesses asymptotics of the $g^a$ type with $\alpha > 1$ can be given a two-fold interpretation. From the standpoint of finiteness of a physical charge at large distances, the $\varphi^4$ theory is inconsistent: the effective charge $g(L)$ turns into infinity at a certain $L_c$ (Landau pole), while for $L < L_c$ the $g(L)$ is undetermined. Considering the field theory as a limiting case of the lattice theories, the $\varphi^4$ theory is "trivial": the physical charge tends to zero for any value of the bare charge.

In recent years, the problems related to the concept of triviality were actively discussed by several researchers (see [30, 31] and references therein). On the one hand, the existing indications of triviality of the $\varphi^4$ theory were emphasized; on the other hand, the $\varphi^4$ theory was declared verified (with a positive result) by numerical modeling on a lattice. Let us briefly discuss this problem as well.

The $\varphi^4$ theory is strictly proven to be trivial in a space with the dimensionality $d > 4$ and nontrivial for $d < 4$ [32, 33]. In the case of $d = 4$, the obtained inequalities were only slightly insufficient for the statement of triviality. This fact was considered by mathematicians as annoying unpleasanty and triviality of the $\varphi^4$ theory was stated as "virtually proved." From the standpoint of physics, this optimism is by no means justified: on the modern level, the aforementioned results for $d \ne 4$ are rather primitive, being simple consequences of renormalizability and a one-loop renormalization group. On the contrary, the situation with $d = 4$ is physically highly complicated and no analytical approaches to solving this problem have been developed so far.

In the author's opinion, the results of numerical experiments on the lattice revealed nothing unexpected. In view of the absence of zeros of the β function, the effective charge $g(L)$ always decreases with the distance. However, the numerical methods cannot answer the question as to whether the "charge zero" does exist, which is explained by limited lattice dimensions. There are many cases of misunderstanding related to the charge normalization: even in the "natural" normalization used in this work, the quadratic law is extended to $g \sim 10$ (see Section 8.3); traditional normalizations extend this interval even greater, for example, up to $g \sim 600$ when the interaction term is written in the form of $g\varphi^4/8$. Therefore, behavior of any quantities is indistinguishable from trivial in a broad range of parameters.

Among old publications, only the paper of Freedman *et al.* [34] is worth of mentioning where it was stated that $g(L)$ uniformly decreases in $g_0$, which is actually indicative of the "charge zero." However, judging by the results, the charge normalization employed in [34] differed by a factor of about 100 (an expression for the action obviously contains a misprint) from that used in this work and all results for finite $g_0$ fell within a region where the quadratic law is operative. Nontrivial results were only obtained for $g_0 = \infty$ by reduction to the Ising model. Although this reduction is apparently possible, there is no method (except for extrapolation) to establish a correspondence between normalization of the field variable in the Ising model and that in the initial $\varphi^4$ theory. This leads to uncertainty in the charge normalization, an allowance for which makes unjustified any conclusions concerning uniform convergence.

[10] This law is more clearly pronounced for the Borel image and is somewhat distorted for the β function as a result of integration in Eq. (8); however, $\beta(g)$ remains downward-convex up to $g \sim 100$.

Now let us turn to the original results of [30, 31]. The main idea was illustrated by the example of a nonideal Bose gas possessing a well-known spectrum of the Bogolyubov type: $\epsilon(k) \sim k$ for small $k$ and $\epsilon(k) \sim k^2$ for $k \longrightarrow \infty$. Let us pass to the "continuum limit" by allowing two characteristic scales of the problem (scattering length and interparticle distance) to tend to zero. If the first value tends to zero rather rapidly, a "quite trivial theory" appears and a quadratic spectrum of the ideal gas is restored. If the limiting transition is performed so as to maintain a certain relationship (ensuring constant sound velocity) between the two scales, a "trivial theory with nontrivial vacuum" appears and the spectrum becomes strictly linear (i.e., strongly different from that of the ideal gas), although no interaction of quasiparticles (phonons) takes place. The latter scenario was suggested for the continuum limit of the $\varphi^4$ theory, stating that it is logically self-consistent.

Even if the last statement is accepted, a question still remains unanswered as to why this limiting transition does physically take place. For a Bose gas of neutral atoms, there is no real possibility of simultaneously changing both the gas density and the scattering length. The situation required for the authors of [30, 31] may take place only in the case of a special long-range interaction, whereby a change in the density affects the Debye screening radius. However, this scenario is not arbitrary and can be predicted based on the initial Hamiltonian.

It was stated [30, 31] that the assumption concerning a nontrivial character of the continuum limit was confirmed by the results of numerical modeling on the lattice. However, this conclusion was based only on a particular interpretation of the "experimental" data, rater than on a direct experimental evidence: the numerical experiments were performed deep in the region of the single-loop law and could not contain any information concerning the triviality. The results, however unusual they might seem, must by explained within the framework of a weak coupling limit.

Triviality of the $\varphi^4$ theory leads to the non-renormalizability of the Higgs spectrum of the Standard Model. This results in violating one of the basic postulates, the principle of renormalizability. Thus, papers [30, 31] were stimulated by the wish to resolve the difficulties. According to the results obtained in this work, no such difficulties were inherent in the system studied.

## 9. CONCLUSION

This paper develops an algorithm for summing divergent series of the perturbation theory with arbitrary values of the coupling constant. Verification on the test examples showed that the algorithm is stable under conditions of strongly restricted information and confirmed reliability of the error estimation. The main physical result of this study consists in restoring the Gell-Mann–Low function of the $\varphi^4$ theory and demonstrating its self-consistency. The latter conclusion agrees with the absence of renormalon singularities established previously [9].

The proposed algorithm can be applied to solving many other problems as well, in particular, to restoring the Gell-Mann–Low functions in quantum electrodynamics and quantum chromodynamics. At present, solving this task is complicated by the absence of calculations of the full-scale Lipatov asymptotics in these theories, although the basis for such calculations is fully prepared [27, 35–39]. Application of the proposed algorithm to the theory of phase transitions may increase the accuracy of calculation of the critical indices by at least two–three orders of magnitude.

## ACKNOWLEDGMENTS

This study was supported by the INTAS foundation (grant no. 99-1070) and by the Russian Foundation for Basic Research (project no. 00-02-17129).

## REFERENCES

1. I. M. Suslov, Pis'ma Zh. Éksp. Teor. Fiz. **71**, 315 (2000) [JETP Lett. **71**, 217 (2000)].
2. L. N. Lipatov, Zh. Éksp. Teor. Fiz. **72**, 411 (1977) [Sov. Phys. JETP **45**, 216 (1977)].
3. *Large Order Behavior of Perturbation Theory*, Ed. by J. C. Le Guillou and J. Zinn-Justin (North-Holland, Amsterdam, 1990).
4. J. Zinn-Justin, Phys. Rep. **70**, 109 (1981).
5. E. B. Bogomolny, V. A. Fateyev, and L. N. Lipatov, Sov. Sci. Rev., Sect. A **2**, 247 (1980).
6. J. C. Le Guillou and J. Zinn-Justin, Phys. Rev. Lett. **39**, 95 (1977); Phys. Rev. B **21**, 3976 (1980).
7. G. A. Baker, Jr., B. G. Nickel, M. S. Green, and D. I. Meiron, Phys. Rev. Lett. **36**, 1351 (1976); Phys. Rev. B **17**, 1365 (1978).
8. J. C. Le Guillou and J. Zinn-Justin, J. Phys. Lett. **46**, L137 (1985); J. Phys. (Paris) **48**, 19 (1987); **50**, 1365 (1989).
9. I. M. Suslov, Zh. Éksp. Teor. Fiz. **116**, 369 (1999) [JETP **89**, 197 (1999)].
10. N. N. Bogoliubov and D. V. Shirkov, *Introduction to the Theory of Quantized Fields* (Nauka, Moscow, 1976; Wiley, New York, 1980).
11. V. S. Popov, V. L. Eletskiĭ, and A. V. Turbiner, Zh. Éksp. Teor. Fiz. **74**, 445 (1978) [Sov. Phys. JETP **47**, 232 (1978)].
12. D. I. Kazakov, O. V. Tarasov, and D. V. Shirkov, Teor. Mat. Fiz. **38**, 15 (1979).
13. Yu. A. Kubyshin, Teor. Mat. Fiz. **58**, 137 (1984).
14. A. N. Sissakian *et al.*, Phys. Lett. B **321**, 381 (1994).

15. A. A. Vladimirov and D. V. Shirkov, Usp. Fiz. Nauk **129**, 407 (1979) [Sov. Phys. Usp. **22**, 860 (1979)].
16. M. V. Sadovskiĭ, Usp. Fiz. Nauk **133**, 223 (1981) [Sov. Phys. Usp. **24**, 96 (1981)].
17. I. M. Suslov, Usp. Fiz. Nauk **168**, 503 (1998) [Phys. Usp. **41**, 441 (1998)].
18. F. M. Dittes, Yu. A. Kubyshin, and O. V. Tarasov, Teor. Mat. Fiz. **37**, 66 (1978).
19. Yu. A. Kubyshin, Teor. Mat. Fiz. **57**, 363 (1983).
20. G. H. Hardy, *Divergent Series* (Clarendon, Oxford, 1949; Inostrannaya Literatura, Moscow, 1951).
21. Yu. V. Sidorov, M. V. Fedoryuk, and M. I. Shabunin, *Lectures on Theory of Functions of Complex Variable* (Nauka, Moscow, 1976), Para. 32.
22. W. H. Press, B. P. Flannery, S. A. Teukolsky, and W. T. Vetterling, *Numerical Recipes* (Cambridge Univ. Press, Cambridge, 1988).
23. C. M. Bender and T. T. Wu, Phys. Rev. **184**, 1231 (1969); Phys. Rev. D **7**, 1620 (1973).
24. J. Cizek and E. R. Vrskay, Int. J. Quantum Chem. **21**, 27 (1982).
25. I. M. Suslov, Zh. Éksp. Teor. Fiz. **117**, 659 (2000) [JETP **90**, 571 (2000)].
26. A. I. Mudrov and K. B. Varnashev, Phys. Rev. E **58**, 5371 (1998).
27. S. V. Faleev and P. G. Silvestrov, Nucl. Phys. B **463**, 489 (1996).
28. S. V. Faleev and P. G. Silvestrov, Phys. Lett. A **197**, 372 (1995).
29. L. N. Lipatov, Zh. Éksp. Teor. Fiz. **71**, 2010 (1976) [Sov. Phys. JETP **44**, 1055 (1976)].
30. M. Consoli and P. M. Stevenson, Z. Phys. C **63**, 427 (1994).
31. A. Agodi, G. Andronico, P. Cea, *et al.*, Mod. Phys. Lett. A **12**, 1011 (1997).
32. J. Frolich, Nucl. Phys. B **200** (FS4), 281 (1982); M. Aizenman, Commun. Math. Soc. **86**, 1 (1982).
33. J. P. Eckmann and R. Epstein, Commun. Math. Phys. **64**, 95 (1979).
34. B. Freedman, P. Smolensky, and D. Weingarten, Phys. Lett. B **113B**, 481 (1982).
35. E. B. Bogomolny and V. A. Fateyev, Phys. Lett. B **71B**, 93 (1977); L. N. Lipatov, A. P. Bukhvostov, and E. I. Malkov, Phys. Rev. D **19**, 2974 (1979).
36. G. Parisi, Phys. Lett. B **66B**, 382 (1977).
37. C. Itzykson, G. Parisi, and J. B. Zuber, Phys. Rev. D **16**, 996 (1977); R. Balian, C. Itzykson, G. Parisi, and J. B. Zuber, Phys. Rev. D **17**, 1041 (1978).
38. E. B. Bogomolny and V. A. Fateyev, Phys. Lett. B **76B**, 210 (1978).
39. I. I. Balitsky, Phys. Lett. B **273**, 282 (1991).